\documentclass[final]{elsarticle}
\usepackage[utf8]{inputenc}
\UseRawInputEncoding
\usepackage{blindtext}
\usepackage{geometry}
\usepackage[nodisplayskipstretch]{setspace}
\usepackage{amsmath}
\usepackage{amssymb}
\usepackage{caption}
\usepackage{graphicx}
\usepackage{latexsym}
\usepackage{times}
\usepackage{lineno}
\usepackage{subcaption,tikz}
\usepackage{calc,float,makecell}
\usepackage{booktabs,array}
\usepackage{epsfig}
\usepackage{placeins}
\usepackage{tabularx}
\usepackage{romannum}
\usepackage{adjustbox}
\usepackage[version=4]{mhchem}
\PassOptionsToPackage{hyphens}{url}
\usepackage[percent]{overpic}
\usepackage[hidelinks]{hyperref} 

 \usepackage{url}
\journal{Atmospheric Pollution Research}
\makeatletter
\def\ps@pprintTitle{%
  \let\@oddhead\@empty
  \let\@evenhead\@empty
  \def\@oddfoot{}%
  \let\@evenfoot\@oddfoot}
\newcommand*{\rom}[1]{\expandafter\@slowromancap\romannumeral #1@}
\makeatother
\graphicspath{{./figures/}}
\biboptions{authoryear}
\begin{document}
\hypersetup{pageanchor=false}
\begin{frontmatter}

\title{On the near-field evolution of a chemically reactive plume behind a ship}

\author[CARESad]{Ramesh Kolluru\corref{correspond}}
\cortext[correspond]{Corresponding author}
\ead{rameshkolluru43@gmail.com}
\author[CAMad]{Savvas Gkantonas}
\author[CARESad]{B Harikrishnan}
\author[CAMad,CARESad]{Epaminondas Mastorakos}
\address[CARESad]{CARES, Cambridge Centre for Advanced Research and Education in Singapore, Singapore}
\address[CAMad]{Department of Engineering, University of Cambridge, Cambridge, UK}

\date{}

\begin{abstract} 
The near-field evolution of chemically reactive pollutants in the exhaust plume of a 270 m crude oil tanker is investigated using a high-fidelity Computational Fluid Dynamics (CFD) framework coupled with a Conditional Moment Closure (CMC) turbulence\--chemistry interaction model and the CBM-IV atmospheric chemistry scheme. Two complementary approaches--a canonical zero-dimensional CMC (0D-CMC) and a spatially resolved Imperfectly Stirred Reactor Network (ISRN)--quantify micromixing effects using information from the CFD flow field.
Micromixing strongly modulates the evolution of NO, NO$_2$, O$_3$, and PAN, explaining spatial discrepancies between detailed CFD\--CMC results and Gaussian plume estimates. While conventional Gaussian formulations overpredict near-field NO by up to $45\%$ within the first 300 m, they are used here as a contextual reference to highlight where turbulence\--chemistry coupling matters. The CFD\--CMC\--ISRN framework therefore provides a physically consistent basis for integrating micromixing effects into next-generation dispersion and air-quality models. This framework bridges the gap between combustion-derived turbulence--chemistry models and environmental dispersion simulations, enabling improved parameterization of subgrid mixing in regulatory contexts.
\end{abstract}


\end{frontmatter}
\hypersetup{pageanchor=true}
\pagenumbering{arabic}

\section{Introduction}

Maritime transport plays a pivotal role in facilitating global trade, with its importance expected to rise in the coming decades~\citep{eyring_2010}. However, it also contributes significantly to air pollution through emissions of nitrogen oxides (NOx), sulphur oxides (SOx), volatile organic compounds (VOCs), and particulate matter (PM), raising concerns among regulatory bodies such as the International Maritime Organization (IMO)~\citep{Lindstad_2015}. These emissions have demonstrable health impacts on coastal populations~\citep{DINATALE2015166} and may contribute to long-term climate forcing. As such, accurately assessing emissions from maritime sources--particularly around densely populated port cities--is essential for the formulation of effective emission inventories and pollution control policies~\citep{Franke_2008}. Although open-sea exposure is limited, understanding near-field reactive chemistry is crucial for quantifying emissions entering the coastal boundary layer and for validating regional models, and the same framework extends directly to anchored or slow-moving vessels in port environments.

One of the core challenges in incorporating ship emissions into atmospheric dispersion models lies in properly characterizing the near-field plume behaviour. Gaussian plume models, commonly used in regulatory applications~\citep{Green_1980, HOLMES_2006, Briggs_1973, yang2020modification}, rely on assumptions of uniform wind and homogeneous turbulence. These assumptions are invalid in the immediate vicinity of ships due to flow separation, turbulence generation, and recirculating wakes. Consequently, the validity of Gaussian models in the near field remains uncertain~\citep{boikos2024validating, mamarikas_assessing_2023}. This uncertainty is particularly relevant for predicting dispersion from exhaust plumes and accidental leaks~\citep{NGUYEN2022104724}.

For chemically reactive species, the situation is even more complex. Rapid dilution and mixing in the near field can significantly influence reaction pathways, especially when chemical timescales are comparable to or shorter than the local turbulent mixing timescales~\citep{mastorakos2003conditional, Garmory_2010}. Therefore, any realistic attempt to model near-field pollutant evolution must consider both macromixing (large-scale turbulent transport) and micromixing (small-scale scalar mixing). The lack of such consideration in conventional models introduces potentially significant predictive errors~\citep{garmory_numerical_2015, kim_investigation_2009}.

Several studies have attempted to address these limitations. \citet{Song_2003} used a Lagrangian photochemical plume model to study NOx chemistry in ship plumes over the North Pacific, showing that plume evolution proceeds through distinct stages. Initial low-reactivity conditions were followed by enhanced ozone and hydroxyl radical production as the plume diluted, resulting in net ozone formation and nitric acid production. \citet{kim_investigation_2009} developed a coupled photochemical--dynamic plume model and highlighted the role of atmospheric stability in modulating primary and secondary pollutant concentrations. However, these studies treated micromixing simplistically, often assuming it to be instantaneous or negligible--an assumption that may not hold in the near field.

The recent SCIPPER Project~\citep{report_D31} reviewed state-of-the-art ship plume models, identifying key uncertainties related to aerosol and gas-phase chemistry, especially in populated areas. However, it also noted that most models devote insufficient attention to the role of turbulent mixing in shaping near-field chemistry. CFD-based approaches, such as those by \citet{murena2018using} and \citet{BOIKOS2024111594}, have improved spatial resolution and captured macromixing effects. More recent work by \citet{boikos2024validating} validated RANS--CFD ship plume models against field measurements in an urban port environment, while \citet{orlandi2024air} presented an integrated forecasting framework combining metocean predictions, ship performance modelling, and pollutant dispersion. 

While these contributions mark important advances, they do not explicitly resolve subgrid-scale turbulence--chemistry interactions or quantify the influence of micromixing in reactive plumes. In contrast, the present work applies the Conditional Moment Closure (CMC) method coupled with a spatially resolved Imperfectly Stirred Reactor Network (ISRN), enabling a physically consistent treatment of micromixing effects alongside finite-rate chemistry. This work quantifies, for the first time, how micromixing-driven turbulence affects reactive pollutant evolution in ship wakes--a process unresolvable by Gaussian models or RANS-only CFD. The Gaussian--CFD comparison remains valuable because it mechanistically links model--data deviations to turbulence--chemistry coupling parameters rather than treating them as generic discrepancies.
The present study addresses this critical need by employing a high-fidelity CFD framework to resolve the wake dynamics of a typical large ship (a 270-meter-long oil tanker) and applying the Conditional Moment Closure (CMC) method~\citep{gkantonas_incompletely_2019} to model the reactive evolution of exhaust species. The CMC framework, originally developed for turbulent combustion, is extended here to atmospheric flows and enables incorporation of both finite-rate chemistry and micromixing effects. The chemical mechanism used is the Carbon Bond Mechanism IV (CBM-IV)~\citep{gery_photochemical_1989}, which accounts for detailed gas-phase photochemistry.

In this context, the mixture fraction $\xi$ is used to track plume dilution, with $\xi = 1$ representing pure exhaust and $\xi = 0$ representing background air. Two complementary CMC-based techniques are used: a canonical zero-dimensional $(0$D$-$CMC$)$ model for sensitivity analysis and a spatially resolved Imperfectly Stirred Reactor Network (ISRN) for detailed plume evolution. This dual approach allows for quantitative evaluation of the influence of micromixing on reactive species such as NOx, SOx, ozone, and VOCs.

The specific objectives of this study are:
\begin{enumerate}
    \item To quantify how micromixing and aerodynamic macromixing jointly shape the chemical evolution of ship exhaust plumes in the near field.
    \item To assess the deviation introduced by ignoring micromixing when using Gaussian or other reduced-order dispersion models.
    \item To discuss implications for parameterising micromixing effects in regulatory and regional air-quality models.
\end{enumerate}

While the focus is on gas-phase pollutants, the results serve as a foundation for extending the framework to aerosol formation (e.g., sulfates, black and brown carbon) in future studies.
\section{Methodology}

\subsection{The modelled ship}

The vessel under investigation is a typical crude oil tanker. The ship has an overall length $L_s ~= ~276$ m, a beam $W_s~ = ~49$ m, and a stack height of $H_s~=~30$ m from the waterline at the design load, corresponding to a gross tonnage of about 83,000 tons and a summer dead weight of 156,000 tons. The vessel has an STX-MAN B\&W 6G70ME-C9.5 2-stroke, 6-cylinder engine with a displacement of about 7,500 l and a rated power of 15.31 MW at 70.8 revolutions per minute operating with marine Heavy Fuel Oil (HFO). 

Here, the ship's speed is taken as 10 knots (5.15 m/s), corresponding to an engine load of 25 \%. We considered a south wind of 3 m / s (light breeze on the Beaufort scale), which is a realistic condition relative to the annual average wind reported by the Singapore Meteorological Service (\url{www.weather.gov.sg}). However, we also include three different scenarios where the wind is against the ship's motion (``headwind''), at right angles to the ship's motion (``crosswind'') and an inclined angle (``Inclined wind''). This has significant consequences on the near-field aerodynamics and how the hull's wake and the infrastructure affect the mixing of the plume.

\subsection{The CFD setup}

CFD simulations were performed with the CONVERGE CFD software version 3.1.9 \citep{richards2016converge}, employing MPI parallelism across approximately 240 processors. The velocity, turbulence, and passive scalar fields were computed using a 3D Reynolds-Averaged Navier-Stokes (RANS) approach with the $k-\omega$ SST turbulence model, incorporating second-order accurate numerical schemes. A density-based solver, coupled with PISO iterations, was employed to solve the 3D flow around the ship structure. Default Flux blending schemes were applied for momentum and turbulence, while default solver settings were utilized for the remaining parameters. Body forces were enabled for all simulations to account for buoyancy effects. 

A fixed box located at the origin of the main stack is equipped with an embedded mesh system comprising eight refinement levels. This embedded mesh creates a finer grid around the main stack, enabling the capture of passive scalar values precisely up to $\xi = 0.95$, even when the actual value at that boundary is $\xi =1$. An adaptive mesh refinement (AMR) strategy is also implemented to accurately resolve the passive scalar variable $\xi$ evolution, which ranges between $1 \times 10^{-5}$ and 1. This AMR approach ensures a minimum of six levels of refinement, thereby guaranteeing proper resolution of the passive scalar dynamics around the main stack and within the wake region. By employing this combined mesh refinement strategy, comprehensive coverage and accuracy are achieved in capturing the behaviour of the passive scalar throughout the relevant areas of interest.
  \begin{figure}[htb!]
      \begin{center}
     \renewcommand{\arraystretch}{3}
     \vspace*{-0.3cm}
        \begin{overpic}[scale = 0.25]{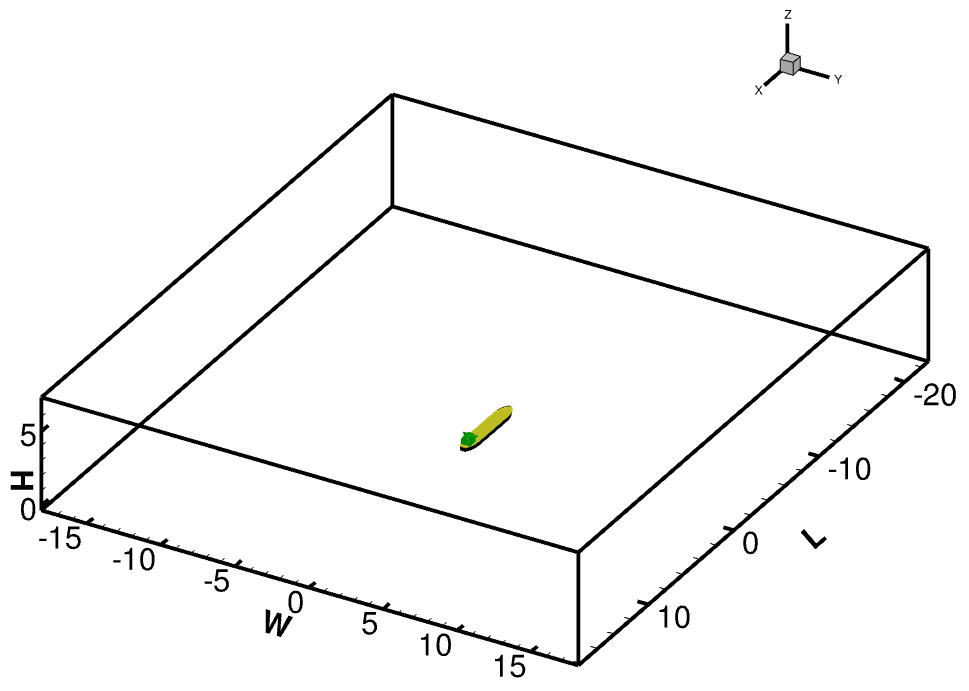}
         \put(70,50){\rotatebox{-15}{\small 1 - Front }}
         \put(65,20){\rotatebox{40}{\small 2  - Left}}
         \put(13,45){\rotatebox{42}{\small 3 - Right}}
         \put(35,35){\rotatebox{15}{\small 5 - Bottom}}        
         \put(65,35){\rotatebox{15}{\small 6 - Top}}         
     \end{overpic} 
      \end{center}
      \caption{Schematic of the computational domain and ship geometry, with all domain dimensions non-dimensionalized by stack height $(H_s = 30 m)$.}
    \label{fig:comp_domain1}
     \end{figure}
Figure~\ref{fig:comp_domain1} illustrates the computational domain employed in this study, along with the boundary conditions described in table~\ref{tab:Boundary_Conditions}. In the case of inclined flow, the ship's geometry is rotated at an angle of $\angle 30^\circ$, and the boundary conditions from the headwind case are applied. Components commonly situated on the top of an oil tanker's deck, such as pipes, cranes, and superstructure elements like antennas, are omitted from the computational domain to simplify the analysis. Prior parametric studies on a similar vessel have demonstrated that this simplification does not significantly impede the analysis's validity~\citep{Park_2016}.
 \begin{table}[H]
            \centering
            \caption{\label{tab:Boundary_Conditions} Boundary conditions for the simulated cases.}
                \begin{adjustbox}{center,max width=\textwidth}
            \begin{tabular}{|c|c|c|c|c|c|c|}
            \hline
         Case/Boundary&Front&Back &Left & Right &Top &Bottom\\
                \hline
            Head Wind&In Flow&Out Flow&Symmetry& symmetry&outflow& no slip\\
                \hline
            Inclined Wind&In Flow&Out Flow&Symmetry& symmetry&outflow& no slip\\
                \hline
            Cross Wind&Symmetry&Symmetry &Out Flow&In Flow&outflow& no slip\\
                \hline
            \end{tabular}
                \end{adjustbox}
            \end{table}

The following notation is used for the geometry of the ship: \(L_s\) denotes the length of the ship, \(W_s\) represents the width of the ship, and \(H_s\) indicates the height of the stack. The dimensions of the computational domain are defined as \(L_c = 7 L_d\), \(W_c = 34 W_s\), and \(H_c = 7 H_d\). These dimensions are further non-dimensionalized using the width of the ship for length and width (\(W_s\)), and the height of the stack (\(H_s\)) for height. The non-dimensional parameters are defined as \(L = \frac{L_d}{W_s}\), \(W = \frac{W_d}{W_s}\), and \(H = \frac{H_d}{H_s}\). In this non-dimensional framework, \(H = 1\) corresponds to the stack height at which pollution or the passive scalar is emitted, and sea level corresponds to \(H = 0\). The dimensions of the computational domain are set to ensure that boundary effects do not influence the flow around the ship region. The chosen length, width, and height for the computation domain are  $L_d = 1860 ~m, W_d = 1645~ m, \text{ and } H_d = 210~ m$ respectively. 

A Cartesian computational grid is employed for grid discretization, with base grid sizes configured as $[dx, dy, dz] = [30, 30, 30]$, $[15, 15, 15]$, and $[10, 10, 10]$. Furthermore, adaptive refinement techniques using a passive scalar are applied to achieve higher resolution in areas of interest. This adaptive refinement strategy results in grid sizes of 35 million, 56 million, and 90 million cells for different grid configurations. The selected grid configuration and adaptive refinement approach facilitate accurate modelling of the flow dynamics around the ship, offering comprehensive insights into the dispersion and behaviour of exhaust gases.

A wind velocity profile is prescribed to represent the oceanic Atmospheric Boundary Layer (ABL) at the computational domain's inlet. This velocity profile is approximated using a power-law given by:
 
 \begin{equation}
 \label{velprofile}
        v  = v_{ref}\left( \frac{ln\left(\frac{z}{z_o}\right)}{ln\left(\frac{z_{ref}}{z_o}\right)} \right)
    \end{equation}
    
\noindent In Eq. \eqref{velprofile}, the reference wind velocity is denoted as $v_{ref}$ and is set to 3 m/s. This reference velocity is measured at a known height above the sea level, given as $z_{ref}$ and set to 10 m. The sea roughness parameter $z_o$ is also assigned a value of 0.001 m, representing typical oceanic conditions. By implementing the power-law velocity profile based on the prescribed parameters, the computational simulations incorporate the influence of the Atmospheric Boundary Layer (ABL) in oceanic conditions at the inlet boundary. This approach enables a more accurate representation of the flow behaviour, capturing the characteristics and effects of the ABL realistically. 

To further characterize the inlet conditions, the turbulent intensity and turbulent integral length scale are prescribed using equations (\eqref{turbinensity}) and (\eqref{lengthscale}), respectively: 
 \begin{equation}
     \label{turbinensity}
        Iv\left(z\right)  = \frac{K_1}{Co\left(z\right)ln\left(\frac{z}{z_o}\right)}
 \end{equation}

 \begin{equation}
     \label{lengthscale}
       L\left(z\right)  = L_t\left(\frac{z}{z_t}\right)^\alpha
\end{equation}
    
\noindent The reference height used in the above equations is $z_t = 200.0 m $, $\alpha = 0.67 + 0.05 ln\left(z_o\right)$, $z_o = 0.001 m$, and $z_o$ is the roughness length.
These equations provide correlations based on guidelines from Eurocode 1 \citep{Eurocode1}, and the specific parameter values are defined as follows:
\begin{enumerate}
    \item  $K_1 = 1.0$: This coefficient calculates turbulent intensity and length scale.
    \item $Co(z) = 1.0$: This represents the turbulence coefficient as a function of height $z$. In this case, it is assumed to be constant and equal to 1.0.
    \item $z_o = 0.001 \, \text{m}$: This refers to the roughness length, representing the characteristic length scale of the surface roughness in the flow domain.
\end{enumerate}
 
\noindent A log-law condition has been used for the velocity boundary conditions on the sea and the ship's surface. A zero-gradient condition has been applied to the outflow planes. 

Finally, a uniform velocity profile was imposed on the exhaust stack under a zero-gradient pressure condition. The exhaust temperature is maintained at 308 K, and $100\%$ humidity was considered, corresponding to the engine conditions at $25\%$  load and a scrubber installation to remove SOx. The equivalent velocity is about 4.6 m/s for an exhaust stack diameter of 1.3 m. The exhaust plume is then replicated using a passive scalar $\xi$, equal to 1 at the exhaust stack and zero in the ambient air background. The buoyancy of the exhaust stack is included and a background temperature of 293 K is assumed. 

The transport equations for the mean and variance of the mixture fraction are calculated as \eqref{MeanXiequation} and \eqref{VarXiequation}
\begin{eqnarray}
\label{MeanXiequation}
    \frac{\partial \bar{\rho} \bar{\xi}}{\partial t} + \frac{\partial \left(\bar{\rho}\bar{u_i}\bar{\xi}\right)}{\partial x_i} = \frac{\partial }{\partial x_i}\left(\bar{\rho} \left( D + D_T\right)\frac{\partial \bar{\xi}}{\partial \bar{x_i}}  \right)
\end{eqnarray}

\begin{equation}
\label{VarXiequation}
    \frac{\partial \bar{\rho} \bar{\xi^{"2}}}{\partial t} + \frac{\partial \left(\bar{\rho}\bar{u_i}\bar{\xi^{"2}}\right)}{\partial x_i} = \frac{\partial}{\partial x_i}\left(\bar{\rho} \left( D + D_T\right)\frac{\partial \bar{\xi^{"2}}}{\partial x_i}  \right) - 2 \bar{\rho}\tilde{N} + 2 \bar{\rho}\left( D + D_T \right) \frac{\partial \bar{\xi}}{\partial x_i}\frac{\partial \bar{\xi}}{\partial x_i}
\end{equation}

with the scalar dissipation rate (SDR) defined as $\tilde{N} = \frac{1}{\bar{\rho}}\left( \rho D \frac{\partial \tilde{\xi}}{\partial x_i} \frac{\partial \tilde{\xi}}{\partial x_i} \right)$. The final two components in \eqref{VarXiequation} are computed, treating them as source terms for the passive scalar. These are then provided as input to the convergence software as a User Defined Function (UDF).

\subsection{Methodology: Conditional Moment Closure (CMC) and Imperfectly Stirred Reactor Network (ISRN)}

The primary aim of this investigation is to analyse how micromixing influences the formation of reactive plumes in the atmosphere. Following an aerodynamics calculation, a turbulent reacting flow theory called Conditional Moment Closure is applied. The evolution of the reacting plume is simulated through the ISRN (Imperfectly Stirred Reactor Network) \citep{gkantonas_0d-cmc_nodate, gkantonas_ISRN_2020} approach. ISRN is a well-accepted method for turbulent non-premixed reacting systems based on the Conditional Moment Closure (CMC) theory. The reacting species concentrations (such as NO, \ce{NO2}, ozone, etc.) as a function of a given value of the passive scalar, the mass fraction of fluid originating from the stack, are quantified by a solution of a transport equation. The underlying fluid mechanical field affects the spatial distribution of the passive scalar, and the way the reacting scalar depends on the passive scalar is determined by the scalar dissipation rate, among other quantities, as shown in the sections below.

Regarding the CMC method, a similar approach to Ref.~\citep{Garmory_2010} is employed, equivalent to the solution of conditional averages of reacting scalars, conditioned on the conserved scalar $\xi$. Given the negligible effect of photolysis on the local temperature, the equations can be decoupled from the flow field calculation. However, the assumption of slight radial variation within the plume and the use of cross-stream conditional averages utilized in Ref.~\citep{Garmory_2010} is relaxed here, and an Imperfectly Stirred Reactor Network is employed instead (see, e.g.,~\citep{gkantonas_incompletely_2019}). The latter method uses a spatially integrated form of the CMC equation that considers spatial effects in all directions. 

\subsubsection{Zero-Dimensional Conditional Moment Closure (0D-CMC)} \label{sec:ODCMC}
The 0D-CMC model is used to establish a reference baseline for evaluating the influence of micromixing on chemically reactive species within a well-mixed environment. This simplified configuration assumes no spatial resolution in physical space and focuses entirely on the evolution of conditional species concentrations in mixture fraction space $\eta$.
The basic governing equations for 0D-CMC modelling is given in Equation \eqref{eqn:0dcmc}:
\begin{equation}
\label{eqn:0dcmc}
\frac{\partial Q_{\alpha}}{\partial t} = \langle N|\eta\rangle \frac{\partial^2 Q_{\alpha}}{\partial \eta^2} + \langle\dot{\omega}|\eta\rangle
\end{equation}
Here :
\begin{itemize}
\item  $Q_{\alpha}$ denotes the chemical species involved in chemical reactions as per the CBM $\rom{4}$ chemical mechanism. 
\item $\langle N|\eta \rangle$ is the conditional scalar dissipation rate, which quantifies micromixing.
\item $\langle \dot{\omega}_\alpha | \eta \rangle$ is the chemical source term, conditioned on the mixture fraction.
\end{itemize}

To model micromixing, the scalar dissipation rate is prescribed using the Amplitude Mapping Closure (AMC) \citep{OBrien_1991}:
\begin{equation}
\langle N | \eta \rangle = N_0 G(\eta), \quad \text{with} \quad G(\eta) = \exp \left( -2 \left[ \text{erf}^{-1}(2\eta - 1) \right]^2 \right)    
\end{equation}
where $N_0$ is a parameter representing average scalar dissipation rate the characteristic micromixing rate at $\eta = 0.5$, offering insights into the predominance of micromixing or chemical reactions.
These equations are solved in the mixture fraction ($\eta$) domain, with boundary conditions outlined in Table \ref{tab:concentrations}. 
This approach allows us to systematically vary $N_0$ to perform a parametric analysis that reveals how different micromixing intensities influence species evolution. 
The simulations are run over a discrete $\eta-\text{grid}$ (51 points) and advanced in time until a steady state is reached (typically at 12 hours). These results serve as qualitative benchmarks for interpreting more spatially resolved ISRN simulations.

\subsubsection{Imperfectly Stirred Reactor Network - ISRN} \label{sec:ISRN}
To capture the spatial variability of chemically reactive species in the ship exhaust plume, Imperfectly Stirred Reactor Network (ISRN) framework is employed. This approach extends the Conditional Moment Closure (CMC) method by embedding a network of virtual reactor cells across the physical domain, each driven by local turbulence and scalar mixing information obtained from the CFD solution. In contrast to the zero-dimensional case, the ISRN method resolves both physical space and mixture fraction space, allowing for a detailed analysis of how turbulence-induced micromixing affects chemical transformations in different regions of the plume.
\begin{sloppypar}
A comprehensive description of the mathematical formulation underlying the ISRN framework is presented in Ref.~\citep{gkantonas_ISRN_2020}.
\end{sloppypar}
The governing equation is given by:
\begin{equation}
\label{eqn:ISRN}
\begin{split}
\frac{1}{V \rho^{**} P^{**}_{\eta}}\left[\sum_{j=1}^{F^{in}}\left(\dot{m}\tilde{P}_{\eta}\right)_j\left[Q_{\alpha}-\left(Q_{\alpha}\right)_j\right] + \sum_{i=1}^{F}\left( \rho D_t \left( \vec{\nabla Q} \right) \tilde{P}_{\eta} \vec{A} \right)_i  \right] 
= \langle \dot{\omega}_{\alpha}|\eta\rangle + N_{\eta}^{**} \frac{d^2 Q_{\alpha}}{d \eta^2}
\end{split}
\end{equation}
\noindent Here,
\begin{itemize}
\item $\rho^{**} = \frac{\int_v \bar{\rho} \tilde{P_{\eta}} dV}{V}$ represents the core volume-averaged mass density, 
\item $P^{**}_{\eta}= \frac{\int_v \bar{\rho} \tilde{P_{\eta}} dV}{V \rho^{**} }$ denotes the mixture fraction probability density function (PDF), 
\item $ N^{**}_{\eta} = \frac{\int_v \bar{\rho} \langle N|\eta \rangle \tilde{P_{\eta}} dV}{V \rho^{**} P^{**}_{\eta}}  ~\refstepcounter{equation}(\theequation)\label{eqn:sdr} $ represents the scalar dissipation rate.
\end{itemize}
 The two terms on the right-hand side of Equation \eqref{eqn:ISRN} signify the chemical reactions and the impact of micromixing. The mixture fraction PDF is estimated from the CFD solution by looking at the mean and variance of the passive scalar and using a $\beta-$function shape \citep{gkantonas_ISRN_2020}. Like the 0D-CMC method, modelling the conditional scalar dissipation rate $N_{\eta}^{**}$ follows a conventional approach based on the AMC method, as described in \citep{gkantonas_ISRN_2020}.

The above equation is a four-dimensional partial differential equation, one for each species in mixture fraction and three physical space coordinates.
An ISRN grid is established in physical space (Supplementary Figure~S10) containing 240 cells, and mixture fraction space is discretized in 101 mesh points. The simulation progresses over time with a time step of $dt = 0.005$ until it reaches a final time of 10000 seconds.

\noindent The solutions from the ISRN code are transferred to the CFD domain as unconditionally averaged values using \eqref{eqn:unconditionalaverage} 

\begin{equation}
\label{eqn:unconditionalaverage}
    Y_{\alpha}^{**} = \int_0^1 \langle{Y_\alpha|\eta}\rangle \tilde{P_{\alpha}}d \eta
\end{equation}

\noindent In this way, the mean species concentrations become available at each CFD cell.

\subsection{Numerical Methodology for OD-CMC and ISRN methods}
The 0D-CMC and ISRN methods described in Sections 2.3.1 and 2.3.2 involve solving governing equations that couple diffusion, chemical reaction, and temporal evolution. These equations are solved using an operator-splitting strategy: the diffusion term is treated with a second-order accurate numerical scheme, while the chemistry is handled through a first-order temporal splitting. Specifically, the diffusion term is integrated using the VODPK solver, followed by the chemical source term integration based on the Carbon Bond Mechanism (CBM-IV).

\subsection{Chemical mechanism and initial conditions}

\begin{sloppypar}
A comprehensive chemical mechanism incorporating photochemical reactions was employed, as described in Ref.~\citep{gery_photochemical_1989}. Originally developed for regional and urban-scale air quality modeling, this mechanism reliably captures the relevant timescales for both stable and radical species, while accounting for their sensitivity to micromixing effects. The species concentrations used as boundary conditions in mixture fraction space for the 0D-CMC and ISRN simulations are summarized in Table~\ref{tab:concentrations}. Here, $\eta = 0$ represents clean background air, and $\eta = 1$ corresponds to the undiluted engine exhaust composition, following the approach outlined in Ref.~\citep{Lee_2017}. For intermediate values of $\eta$, the initial condition is linearly interpolated between the endpoints, consistent with standard practice in turbulent non-premixed mixture fraction modeling. The exhaust concentrations $(\eta = 1)$ are derived from engine certification data under the $E3$ test cycle, in accordance with MARPOL $73/78$ Annex VI guidelines for marine engines. Background concentrations $(\eta = 0)$ are obtained from ambient measurements reported in Ref.~\citep{kim_investigation_2009}.
\end{sloppypar}

\begin{table}[H]
\centering
\caption{\label{tab:concentrations}Initial species boundary conditions at the stack exhaust and background air \citep{kim_investigation_2009}.}
\scalebox{0.9}{
\begin{tabular}{ |c|c|c|}
\hline
Species & Concentration in ppb &  Concentration in ppb \\
\hline
&   $@$ Exhaust & $@$ Background\\
\hline
Nitric Oxide (NO)&111114.7873&0.182\\
\hline
Nitrogen Di Oxide (NO2)&11111.48&0.018\\
\hline
Ethene$\left(ETH\right)$&1.871& 0.35\\
\hline
Xylene$\left(XYL\right)$&0.220& 0.35\\
\hline
Toluene (TOL)&0.538& 0.35\\
\hline
Carbon Monoxide (CO) &58.016& 130.0\\
\hline
Methane (CH4)&68.353& 0.0\\
\hline
Sulpher dioxide (SO2)&43.974&0.0\\
\hline 
PAN & 0.0 & 0.135\\
\hline
Ozone O3 & 0.0 & 20.3\\
\hline
\end{tabular}}
\end{table}


\section{Results and Discussion}
\subsection{Main flow features}
\begin{figure}[htb!]
     \begin{center}
      \vspace*{0cm}
    \renewcommand{\arraystretch}{3}
 \hspace*{-2cm}
    \begin{tabular}{ccc}     
     HeadWind  & Inclined Wind & Cross Wind  \\
    \begin{overpic}[scale = 0.15]{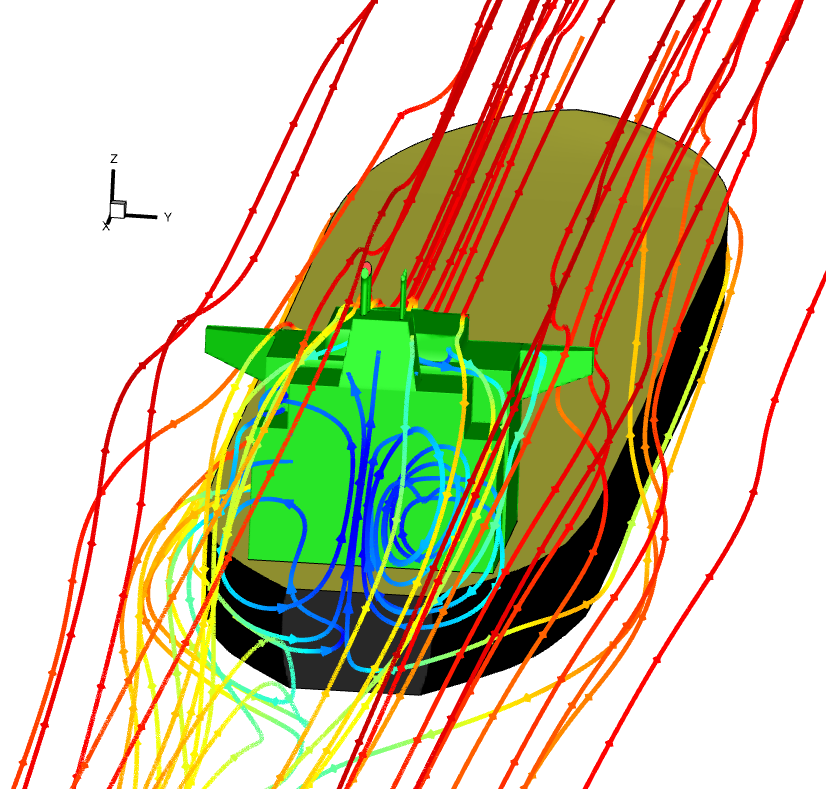}
     \end{overpic} &
         \begin{overpic}[scale = 0.15]{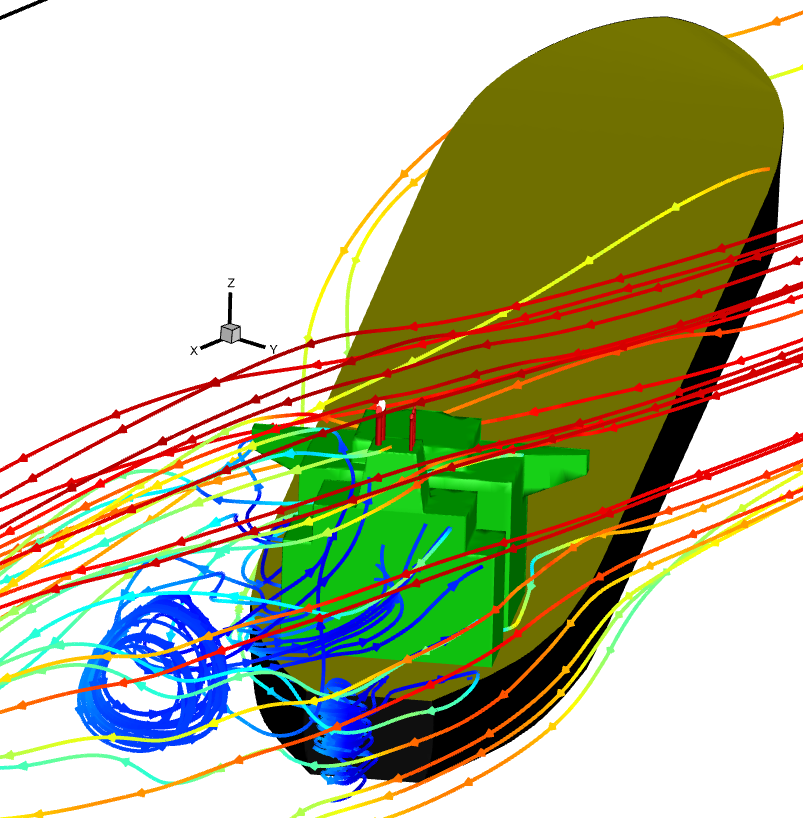}
     \end{overpic}&
         \begin{overpic}[scale = 0.02]{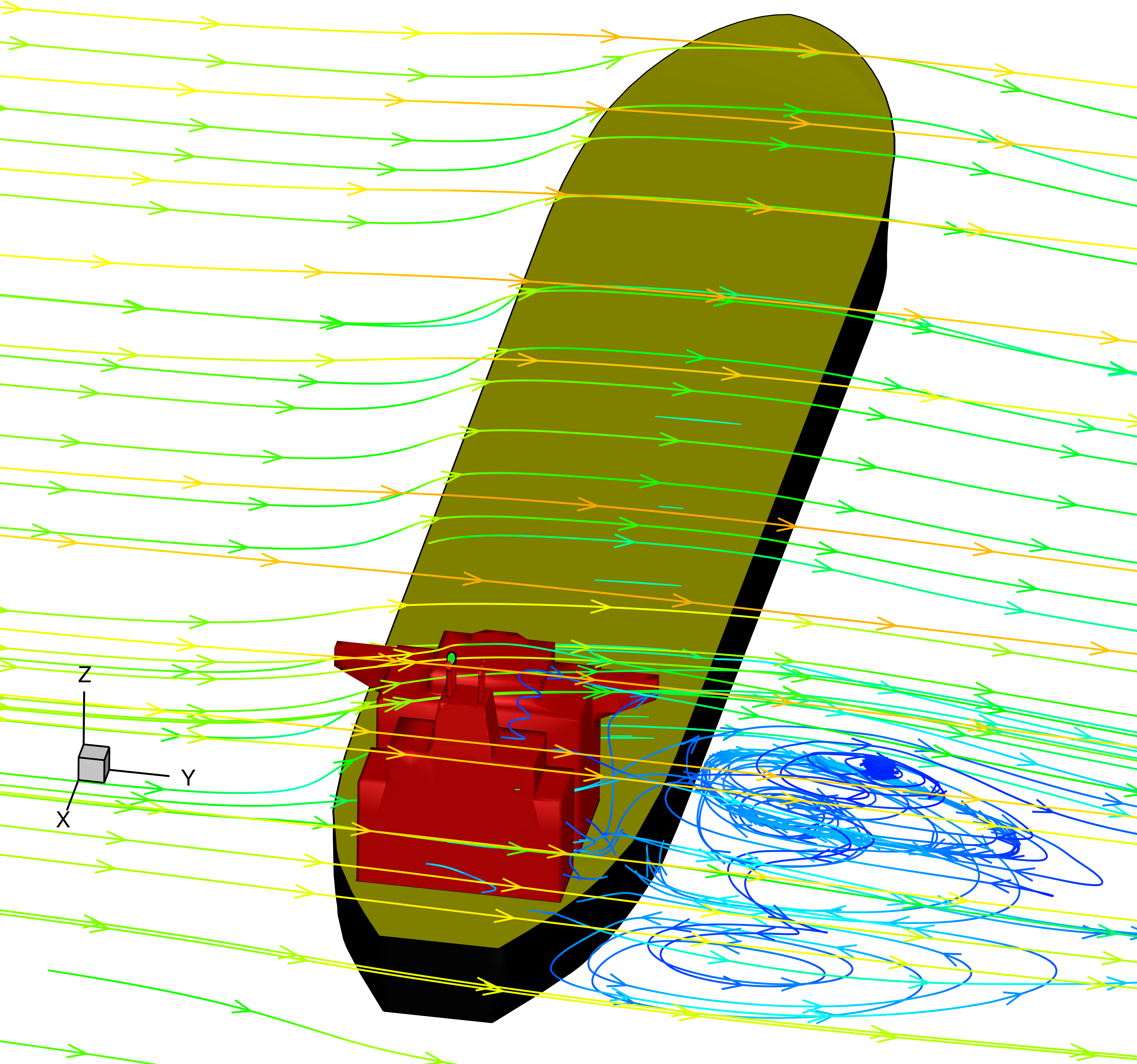}
     \end{overpic}
    \end{tabular}     
     \end{center}
     \caption{Visualization of mean airflow streamlines in the ship's wake for headwind, inclined wind, and crosswind cases. Streamlines are color-coded by longitudinal velocity, illustrating wake extent and turbulence generation behind the superstructure}
      \label{fig:StreamLines3D}
    \end{figure}

Figure~\ref{fig:StreamLines3D} provides a visualization of the mean airflow over the ship through the representation of streamlines. These streamlines have been colour-coded according to the $\bar{u}$ velocity component. Under the investigated conditions, the air wake behind the ship's superstructure extends approximately 1.5 ship widths in the direction of the wind. The wake region exhibits increased turbulence resulting from recirculation and vortical structures.
    
Supplementary Figures~S1--S3 illustrate the isosurfaces and cross-sections of the passive scalar $\xi$ for headwind, inclined wind, and crosswind cases. These isosurfaces use a threshold of $\xi = 10^{-5}$ on a logarithmic scale to highlight highly diluted exhaust mixed with background air. Photolysis reactions and the chemical evolution of reactants occur within finite $\xi$ values where micromixing can be significant~\citep{Garmory_2010}, and plume inhomogeneities are therefore expected~\citep{von_glasow_modeling_2003}. Headwind and crosswind plumes remain largely symmetric, whereas inclined winds yield asymmetry because of macromixing around the hull and superstructure.

\begin{figure}[H]
    \centering
    \begin{adjustbox}{center,max width=\textwidth}
    \begin{tabular}{cc}
        \begin{overpic}[scale=0.25]{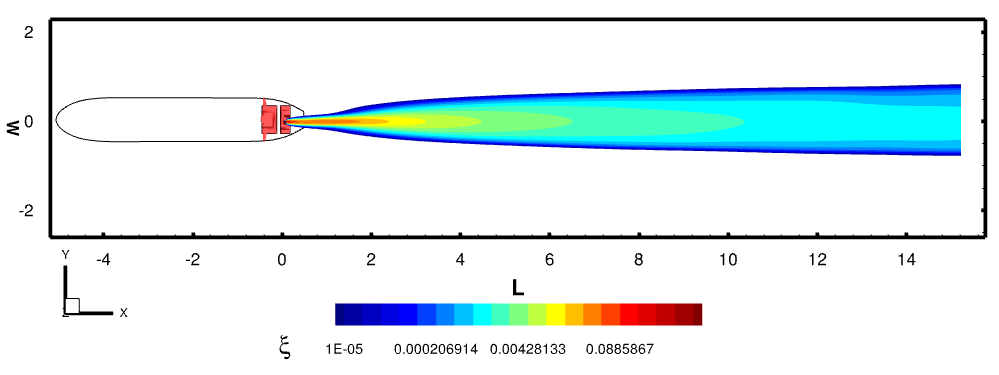}\end{overpic} &
        \begin{overpic}[scale=0.2]{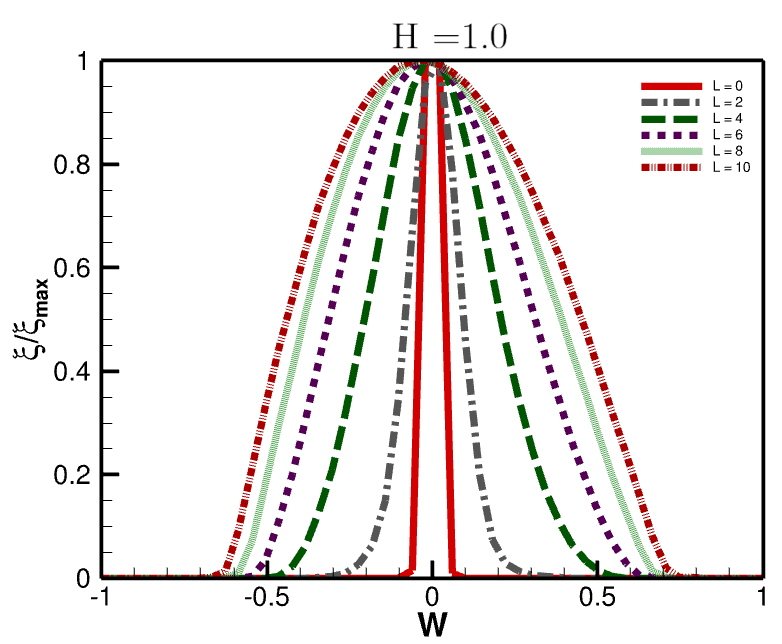}\end{overpic}\\
        \begin{overpic}[scale=0.275]{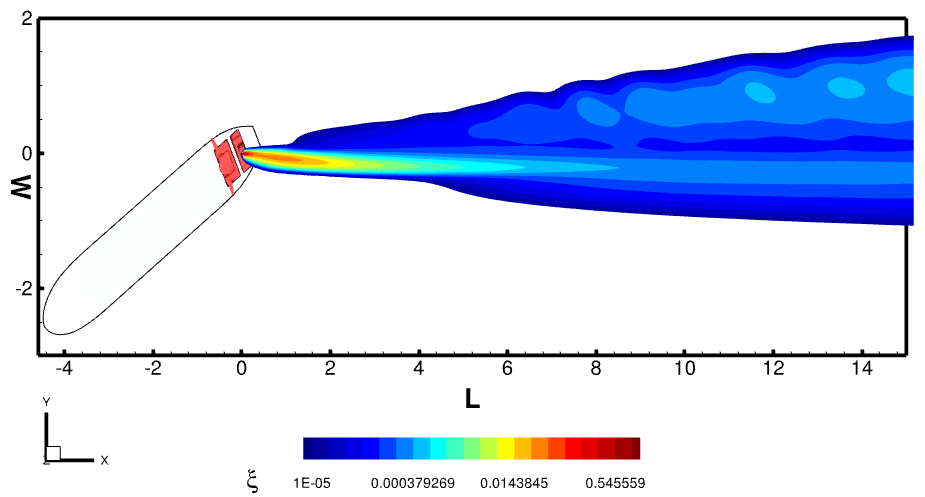}\end{overpic} &
        \begin{overpic}[scale=0.2]{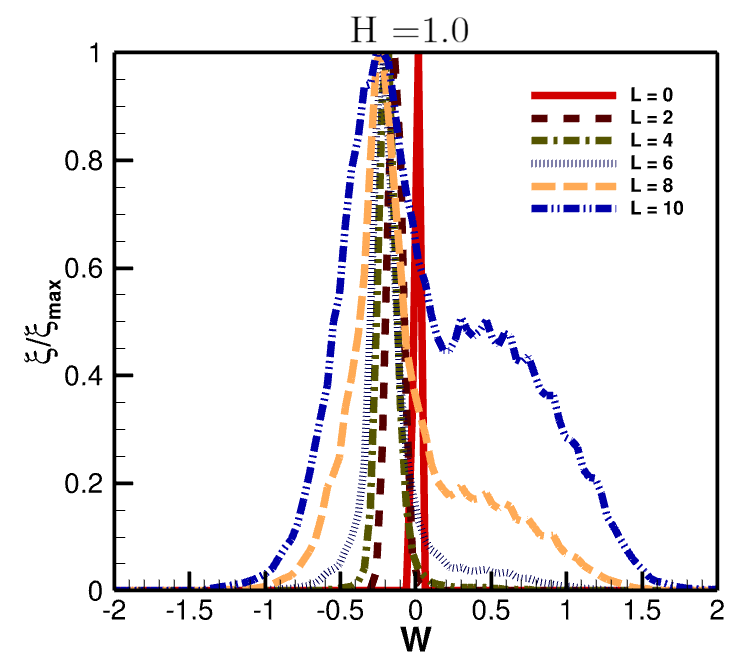}\end{overpic}\\
        \begin{overpic}[scale=0.35]{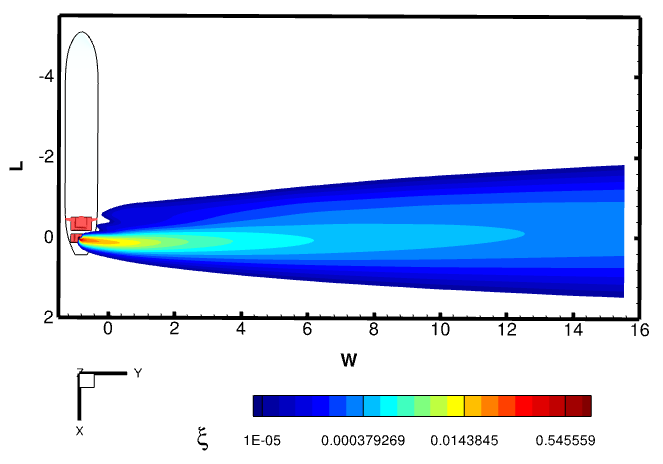}\end{overpic} &
        \begin{overpic}[scale=0.25]{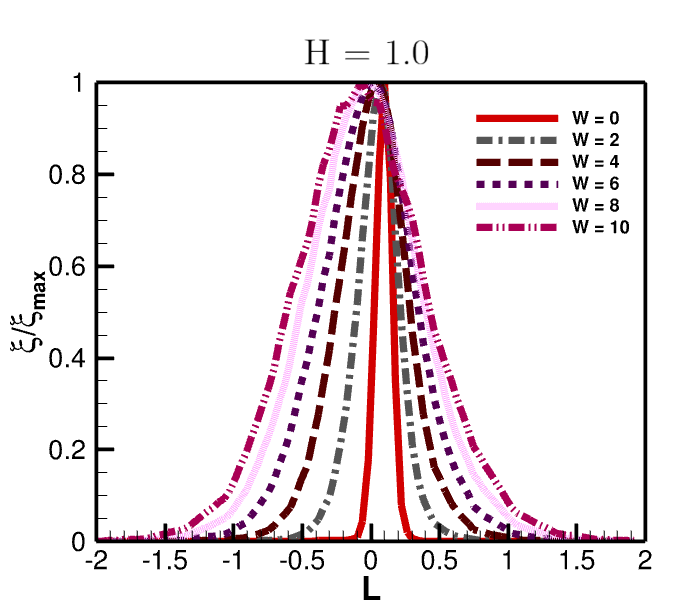}\end{overpic}
    \end{tabular}
    \end{adjustbox}
    \caption{Passive scalar distribution in the $x$--$y$ plane at $z/H_{st}=1$ with horizontal line plots at several downwind distances for headwind, inclined wind, and crosswind cases.}
    \label{fig:xi_z1_main}
\end{figure}

Supplementary Figure~S4 provides the passive scalar concentration contours on $x$--$z$ (or $y$--$z$) planes together with vertical line profiles, documenting how wake confinement shapes dilution across wind cases. Figure~\ref{fig:xi_z1_main} shows the $x$--$y$ plane at $z/H_{st}=1$ with horizontal line plots at several downwind distances for headwind, inclined wind, and crosswind cases, highlighting non-Gaussian dilution patterns one stack height above deck.

\begin{figure}[H]
    \centering
    \begin{adjustbox}{center,max width=\textwidth}
    \begin{tabular}{ccc}
    Head Wind Case & Inclined Wind at  $30^\circ$ & Cross Wind\\
          \begin{overpic}[scale = 0.0275]{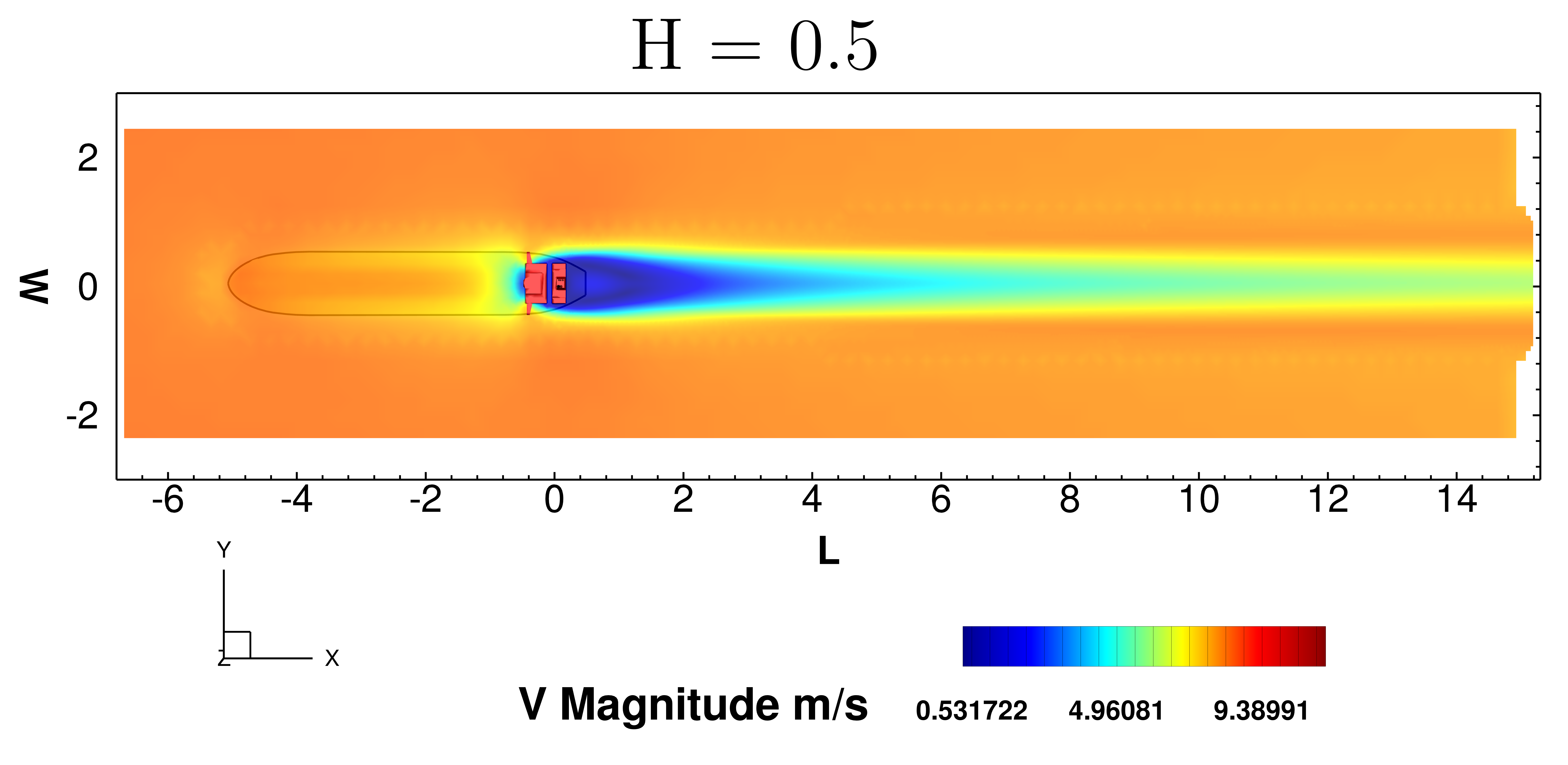}
      \end{overpic} &
         \begin{overpic}[scale = 0.021]{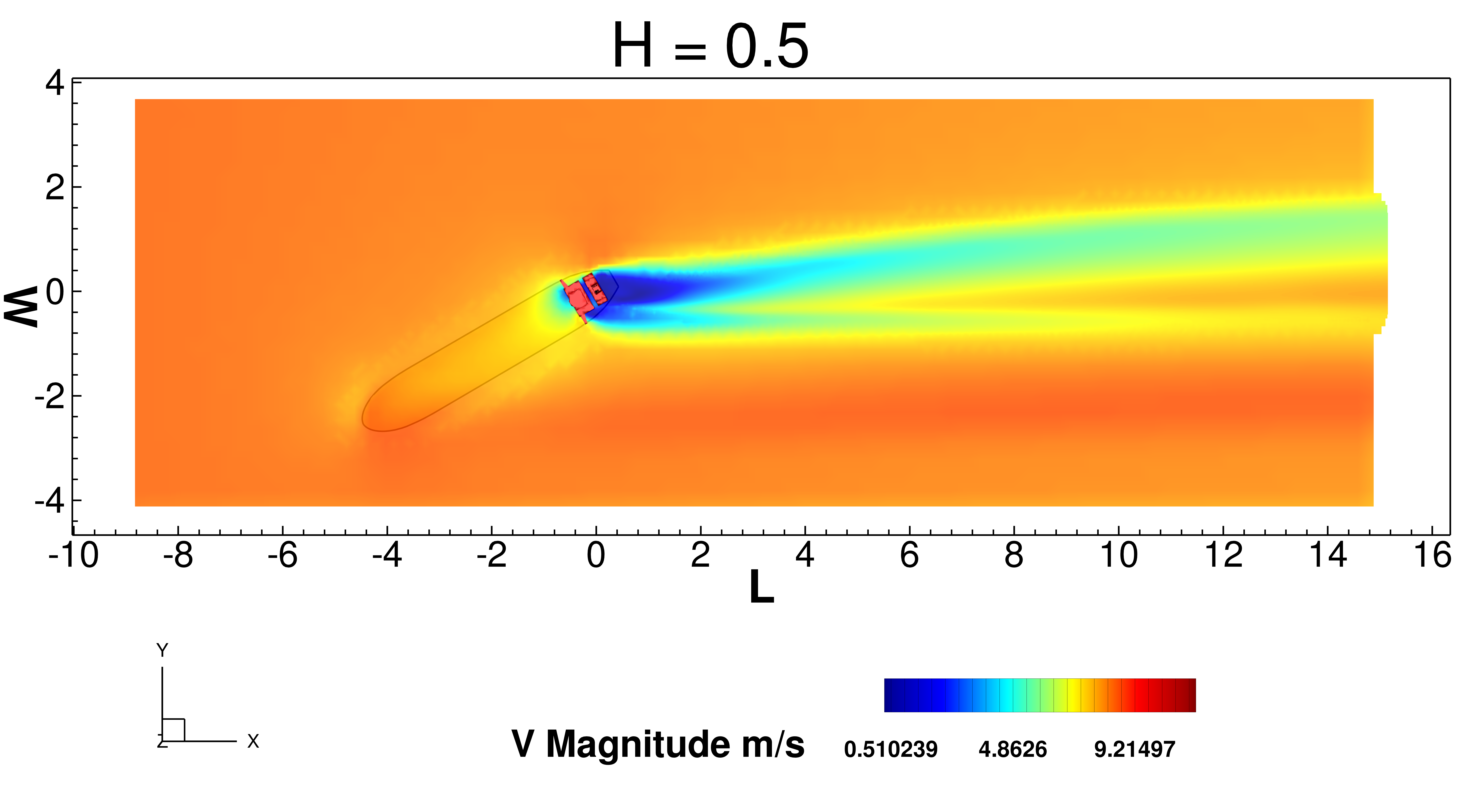}
     \end{overpic}&
         \begin{overpic}[scale = 0.02]{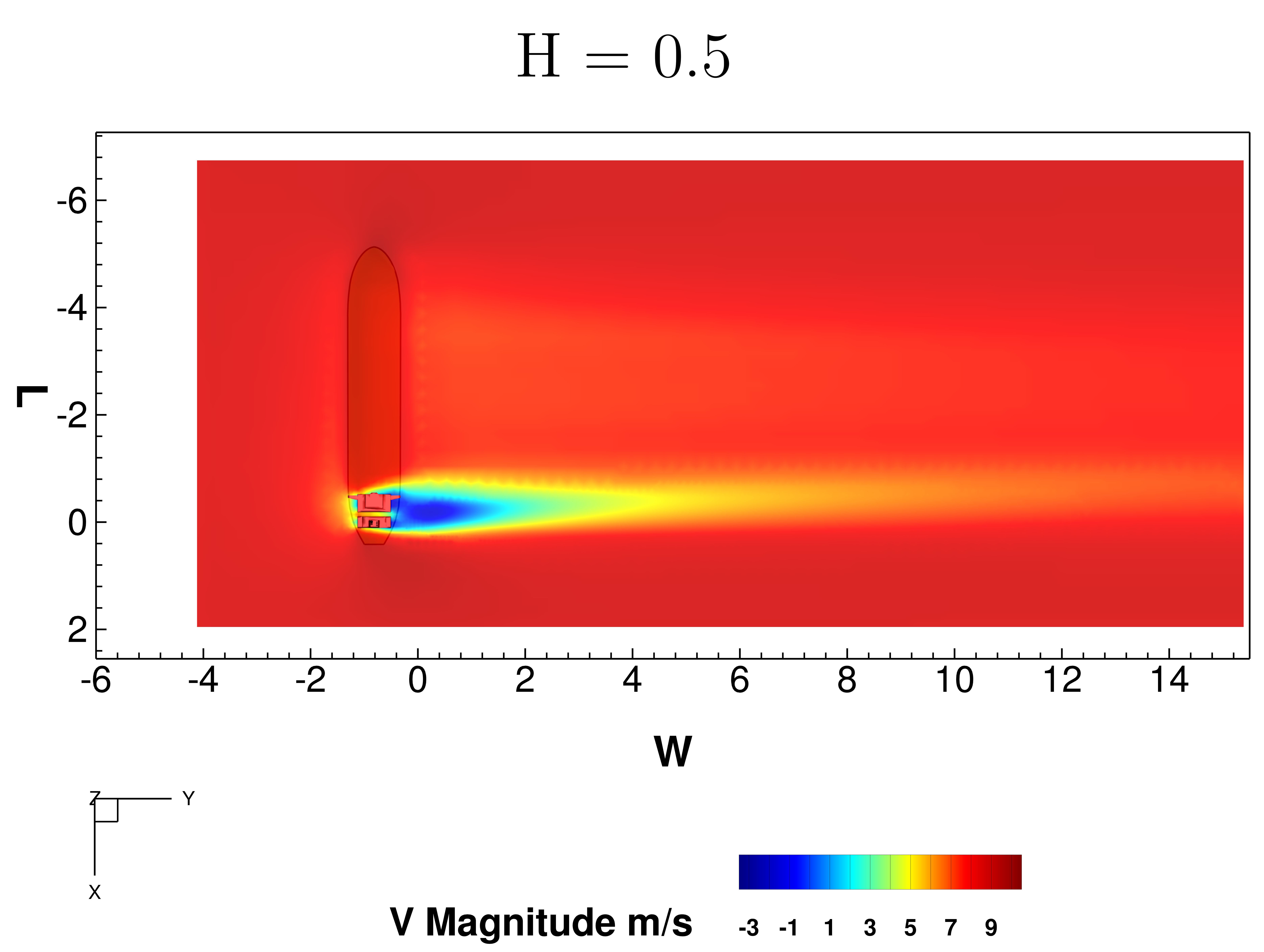}
     \end{overpic}\\
\end{tabular}
    \end{adjustbox}
    \caption{Contours of velocity magnitude at $H = 0.5$ above sea level for headwind, inclined wind, and crosswind cases. Additional contours at $H = 1.0$ are provided in the Supplementary Material.}
     \label{fig:velocitycontours}
    \end{figure}

Figure~\ref{fig:velocitycontours} provides visual representations of the velocity magnitude contours at $H =0.5$ for all simulated cases. These contours reveal essential flow field characteristics in the wake region behind the superstructure. The contours are asymmetric in the inclined wind cases. This asymmetry is consistently observed along the height of the ship, suggesting that the shape and geometry of the superstructure influence the flow patterns. Supplementary Figure~S9 shows the $H=1.0$ velocity slice and confirms that the air wake of the superstructure drives downward motion toward the sea surface under the investigated conditions. This downwash is particularly prominent in the inclined wind and crosswind cases, where the inclination of the wind further enhances this effect. The downward movement of air can have significant implications, such as the entrainment of exhaust gas into the lower regions of the flow. The downwash can trap the plume and bring it downwards on deck and towards the sea level. This is important in the case of toxicity emissions and may also act to move the effective emission point in the case of modeling where the emission is treated as a point source \citep{Pan_2021}.

The superstructure aerodynamics appear dominant within the first meters of the plume, as evident from Figure~\ref{fig:velocitycontours}. Considering different operating scenarios and superstructure geometry, where the exhaust is emitted outside the boundary layer, the gas momentum, buoyancy, and entrainment by ambient air may have a more significant effect, and the exhaust gas would be emitted more smoothly~\citep{Park_2016}. It would, therefore, be interesting to assess the near-field pollution level for different conditions depending on the entrapment of exhaust gas within the ship's air wake. Supplementary Figures~S4 and S5 present the normalized $\xi$ fields and line plots across planes and downwind distances, underscoring the wake-confinement and downwash patterns most relevant to receptor exposure near ports and coastal cities. Integration of such micromixing-informed frameworks into port-scale modeling systems (e.g., EPISODE--CityChem or SCIPPER datasets) can improve predictive fidelity.

In the headwind and crosswind cases, as we move farther from the wake region, the spread of the passive scalar approximates a Gaussian behaviour in the $x-y$ plane (Supplementary Figure~S5). In the inclined wind case, the passive scalar deviates from Gaussian behaviour because a secondary stream through the superstructure channel alters entrainment (Figure~\ref{fig:velocitycontours}). The inclined wind introduces additional flow patterns, leading to non-uniform dispersion. The sea surface further perturbs Gaussian behaviour across all scenarios, especially within the wake region where the plume is trapped (Supplementary Figures~S4 and S5).

\subsection{Comparison with Gaussian Plume Model}
The Gaussian plume model is a prevalent mathematical tool for forecasting the dispersion of air pollutants within the atmosphere. This model is adept at estimating the dissemination of contaminants emitted from specific point sources. Operating under stable atmospheric conditions, the Gaussian plume model posits that pollutants disperse akin to the spread of a passive scalar in a turbulent flow. This model integrates crucial parameters, including source emission rate, stack height, wind speed, wind direction, atmospheric stability, and dispersion coefficients in horizontal ($\sigma_y$) and vertical ($\sigma_z$) dimensions.

In examining the release of a passive scalar from the primary stack, pollutant dispersion is closely influenced by atmospheric conditions. For a comparative analysis of the passive scalar distribution derived from CFD simulations, an emission rate of $Q = 18.3588 \frac{kg}{s}$ is used, reflecting the mass flow rate from the ship's main stack. This stack, positioned approximately 30 meters above the surface, encounters a wind speed of $9.12 \frac{m}{s}$, measured at the height of the stack within the Atmospheric Boundary Layer (ABL). For Gaussian plume calculations, a source density of $\rho_{source} = 1.1657 \frac{kg}{m^3}$ is applied. Dispersion coefficients, calculated using equations \eqref{eqn:Briggs_urban_D_Parameters}, \eqref{eqn:Briggs_open_D_Parameters}, and \eqref{eqn:Pasquill_D_Parameters}, are refined under neutral, stable conditions based on the recommendations by Briggs \citep{Briggs_1973} and further improved by Pasquill \citep{pasquill1979atmospheric}, enhancing the model's accuracy.
\begin{eqnarray}
\begin{aligned}
\label{eqn:Briggs_urban_D_Parameters}
    \sigma_y = 0.16x\left(1+0.0004x\right)^{(-0.5)}\\
    \sigma_z = 0.14x\left(1+0.0003x\right)^{(-0.5)}     
\end{aligned}
\end{eqnarray}
\begin{eqnarray}
\begin{aligned}
\label{eqn:Briggs_open_D_Parameters}
        \sigma_y = 0.08\left(1+0.0001x\right)^{(-0.5)} \\
        \sigma_z = 0.06x\left(1+0.0015x\right)^{(-0.5)} 
\end{aligned}
\end{eqnarray}
\begin{eqnarray}
\begin{aligned}
\label{eqn:Pasquill_D_Parameters}
        a_y = -2.555;  b_y = 1.0423; c_y = -0.0087;\\
        a_z =  -3.186; b_z = 1.1737; c_z = -0.0316;\\
        \sigma_y = e^{(a_y + b_ylog(x) + c_ylog(x)^2)}\\
        \sigma_z = e^{(a_z + b_zlog(x) + c_zlog(x)^2)}
\end{aligned}        
\end{eqnarray}
Supplementary Figure~S1 presents 2D plume contours at \( z/H_{\text{st}} = 1 \) for the four Gaussian dispersion formulations (Briggs urban/open, Pasquill, AERMOD). Supplementary Figure~S2 compares Gaussian plume and CFD profiles at selected downwind locations for headwind, inclined-wind, and crosswind cases. These panels show that Gaussian models under-represent downwash-induced spread aloft and can overpredict concentrations in crosswind scenarios; AERMOD and Pasquill track the trends better than the Briggs variants. The discrepancies underscore the need to consider wake-driven mixing when using reduced-order dispersion models.

\subsection{Evolution of reacting scalars}

\subsubsection{Reference results from 0D-CMC}
\begin{figure}[htb!]
    \centering
     \hspace*{-1.0cm}  
    \begin{tabular}{ccc}
    \includegraphics[scale=0.35]{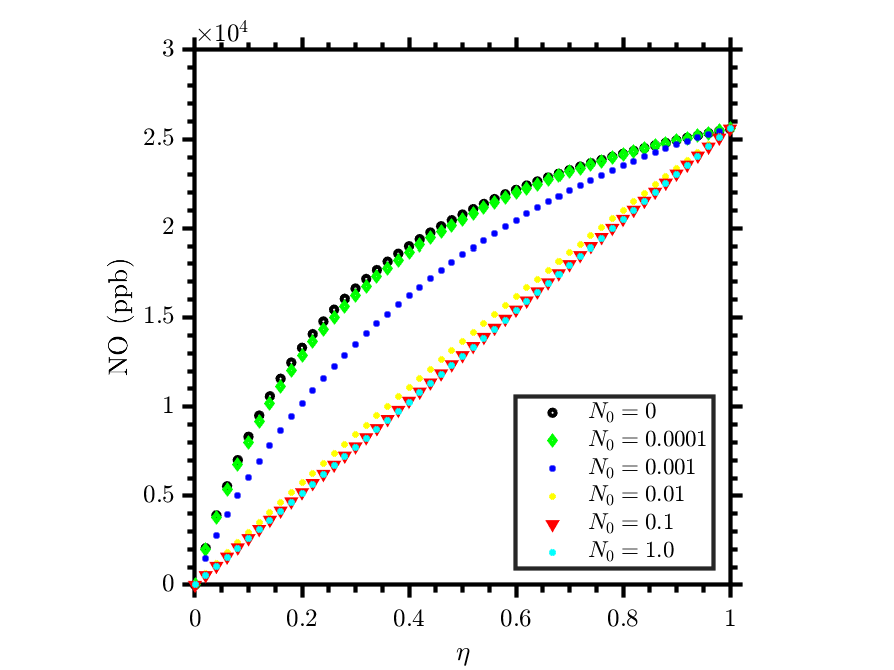}
       & 
      \includegraphics[scale=0.35]{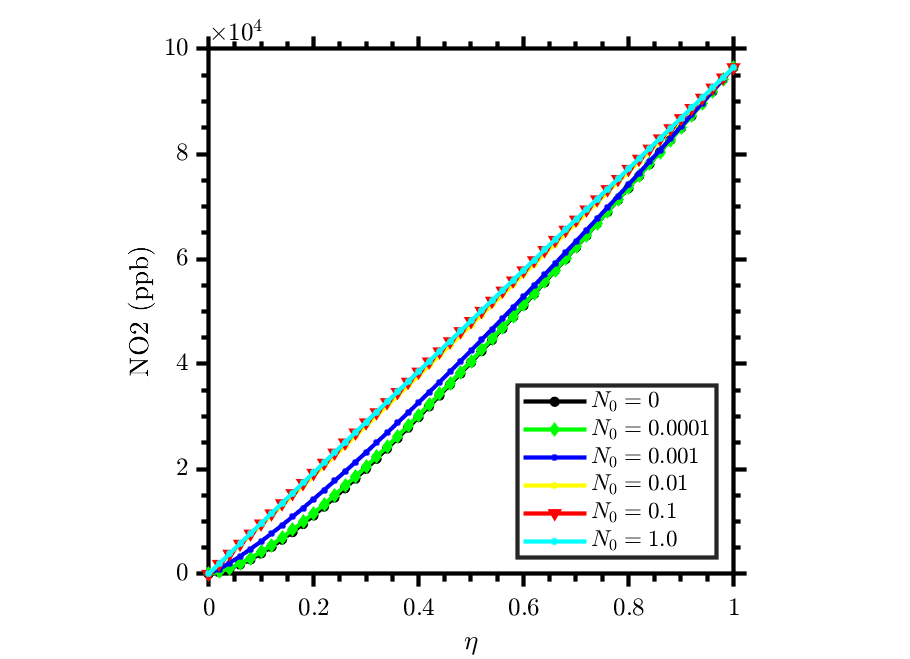}
        &
    \includegraphics[scale=0.35]{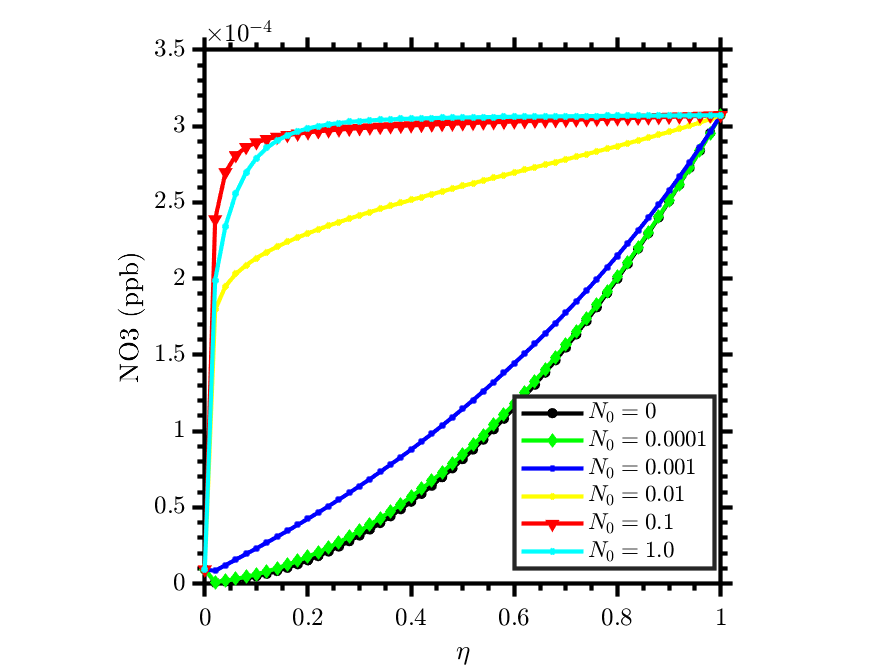}\\
    \end{tabular}
    \caption{Impact of the micromixing rate (quantified by the indicated $N_0$) on nitrogen oxides concentrations at steady-state from the 0D-CMC simulations.}
    \label{fig:cmc0d_atmo_sdrs_1}
\end{figure}

Figure~\ref{fig:cmc0d_atmo_sdrs_1} and Supplementary Figure~S6 show how the micromixing rate shapes different chemical species in mixture-fraction space, spanning the transition from the theoretical limit of $N_0 = 0$ (no mixing between particles) to higher $N_0$ values. Micromixing acts as a strong driver, producing qualitative and quantitative differences in the evolution of $\text{NO}_2$, NO, $\text{O}_3$, PAN, and $\text{HNO}_3$. For sulphur oxides (SO$_2$ and SO$_4$), the micromixing impact is more modest. Encountering realistic $N_0$ values in turbulent wakes therefore implies measurable deviations from simulations that neglect micromixing.

\subsubsection{Results from the ISRN}

Results obtained from the 0D-CMC studies, as illustrated in Figure \ref{fig:cmc0d_atmo_sdrs_1} and Supplementary Figure~S6, suggest that the impact of scalar dissipation rate is minimal at lower values of $N_0$. However, as the scalar dissipation rate (SDR) increases, the effect of micromixing becomes more pronounced in the evolution of specific species such as NO3, SO4, and others. In the real flow, the SDR changes in space, and this is taken into account in the ISRN model that uses different SDR in different locations in space. An approximate assessment of the effect of SDR in the ISRN results can be achieved by setting $N_{\eta}^{**} = 0$ in equation \eqref{eqn:ISRN} and compare it with the naturally evolving $N_{\eta}^{**}$, which incorporates micromixing effects. 

\begin{figure}[H]
        \centering
        \hspace*{-1cm}
        \begin{tabular}{ccc}
        \includegraphics[scale=0.35]{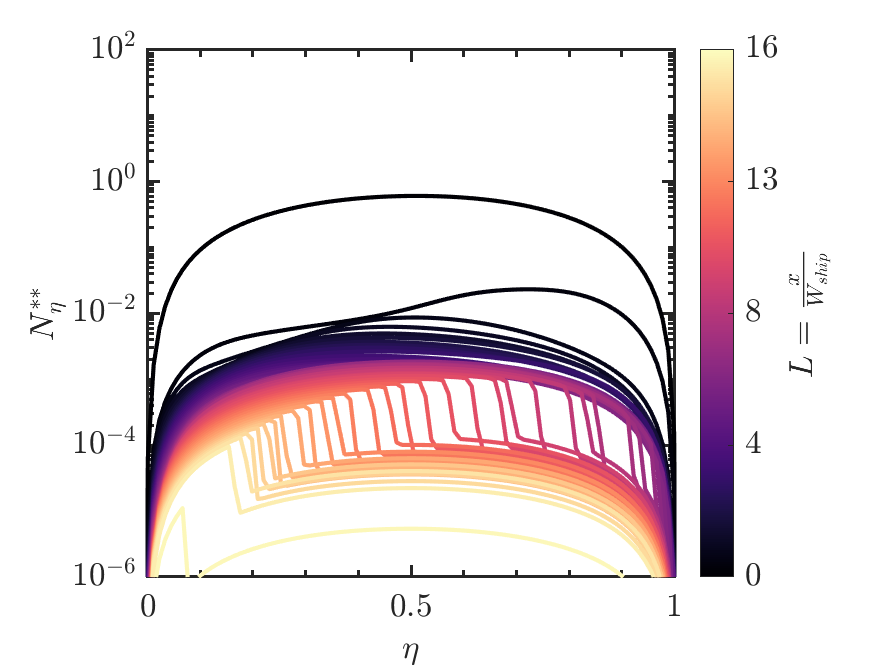} &
            \includegraphics[scale=0.35]{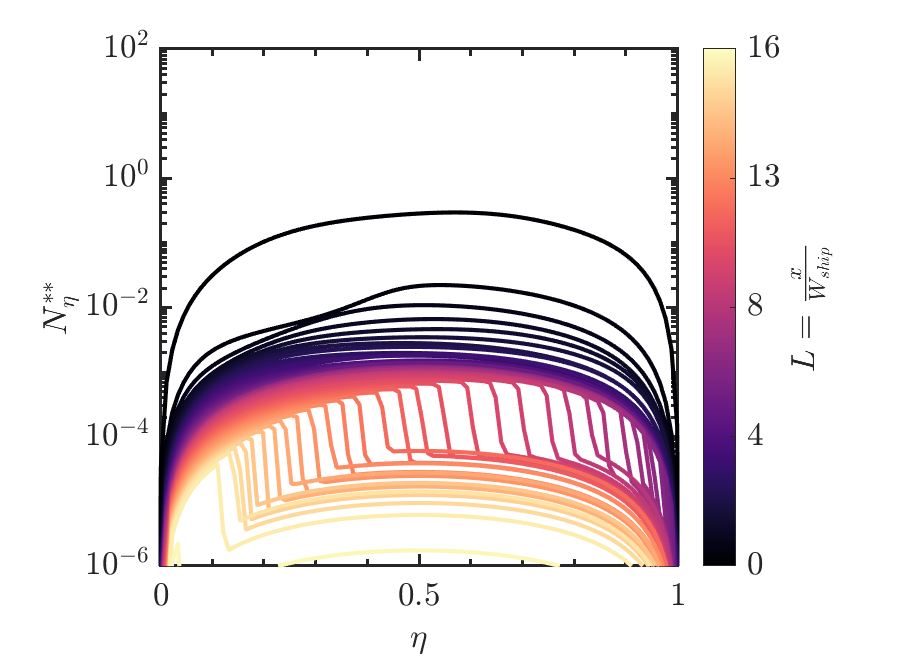}&
            \includegraphics[scale=0.35]{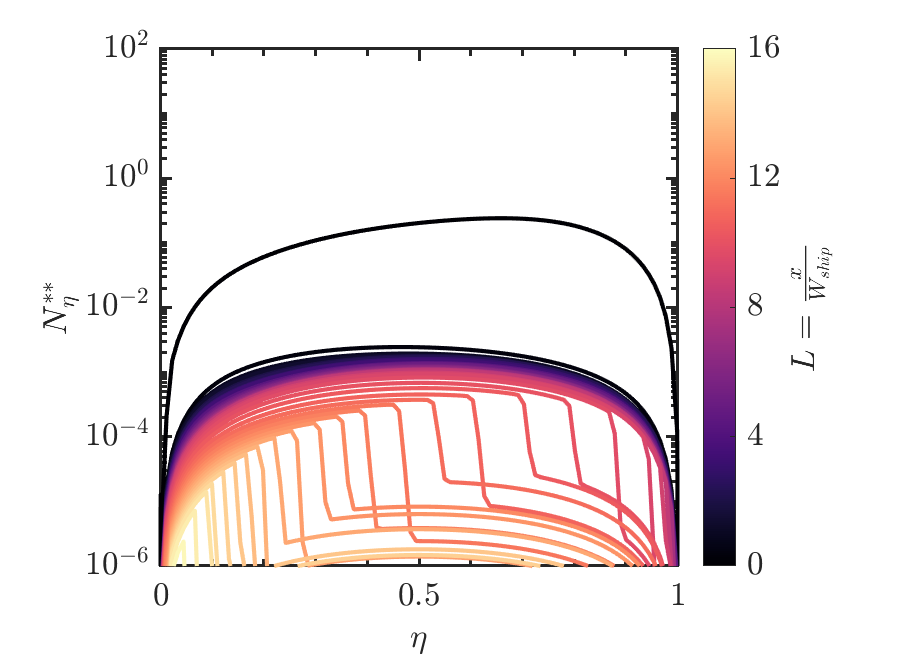}\\
            Head Wind & Inclined Wind & Cross Wind
        \end{tabular}
        \caption{Spatially-integrated scalar dissipation rates $ N_{\eta}^{**} $ for headwind, inclined wind, and crosswind cases in the ship wake region.}
        \label{fig:ThreeCases_sdr}
\end{figure}

Figure \ref{fig:ThreeCases_sdr} illustrates the evolving scalar dissipation rates for different wind scenarios (Head Wind, Inclined Wind, and Cross Wind) in mixture fraction space. Remarkably, the naturally evolving scalar dissipation rate spans from $\left[10^{-6}, 0.2\right]$ within mixture fraction space, determined through the mean and variance of the passive scalar obtained from CFD calculations. Figures \ref{fig:ThreeCases_sdr} reveal that although SDR values decrease as the flow evolves, their significance becomes notably evident in the progression of species such as NO, SO4, HNO3, PAN, and aldehydes as the flow advances. 

The micromixing effect is therefore quantified through the scalar dissipation rate, which ranges from $10^{-6}$--$0.2\,\text{s}^{-1}$ across the wake (Figure~\ref{fig:ThreeCases_sdr}). Sensitivity analysis shows that NO reductions of approximately 10--20\% and PAN increases of roughly 15\% occur for high SDR, confirming the quantitative role of micromixing on fast-reacting species.

Supplementary Figures~S7 and S8 show the conditionally averaged species in mixture-fraction space from the ISRN calculations for the headwind case. The panels demonstrate that species such as NO, NO$_2$, SO$_4$, HNO$_3$, PAN, ALD, and O$_3$ are sensitive to micromixing, in agreement with the 0D-CMC trends. Other species exhibit only minor changes.

In Supplementary Figure~S7 (non-zero SDR), micromixing effects are strongest for NO close to the wake where turbulence is highest (curves coloured by larger unconditional means), and they taper off downstream. In contrast, Supplementary Figure~S8 (SDR$=0$) lacks this smoothing and redistribution, highlighting the role of micromixing in shaping conditional profiles. The comparison indicates distinct behaviours by species: for NO the micromixing impact is most pronounced near the source and weakens with distance, whereas for SO$_4$ and HNO$_3$ the relative differences persist farther from the wake as these species form and evolve. While the chemistry of nitrogen oxides (NO$_x$) governs oxidants including OH, O$_3$, and NO$_3$, the direct imprint of micromixing on OH, O$_3$, and NO$_3$ is comparatively modest.

\begin{figure}[H]
     \renewcommand{\arraystretch}{-0.5}
     \centering
     \begin{adjustbox}{center,max width=\textwidth}
\begin{tabular}{cc}
        \begin{overpic}[scale=0.25]{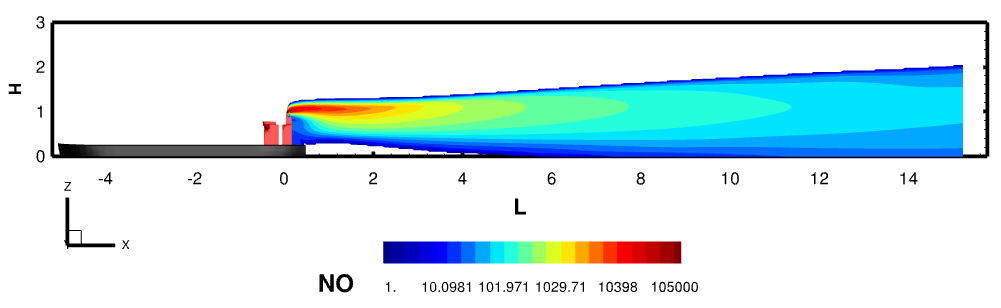}
     \end{overpic} &
         \begin{overpic}[scale=0.2]{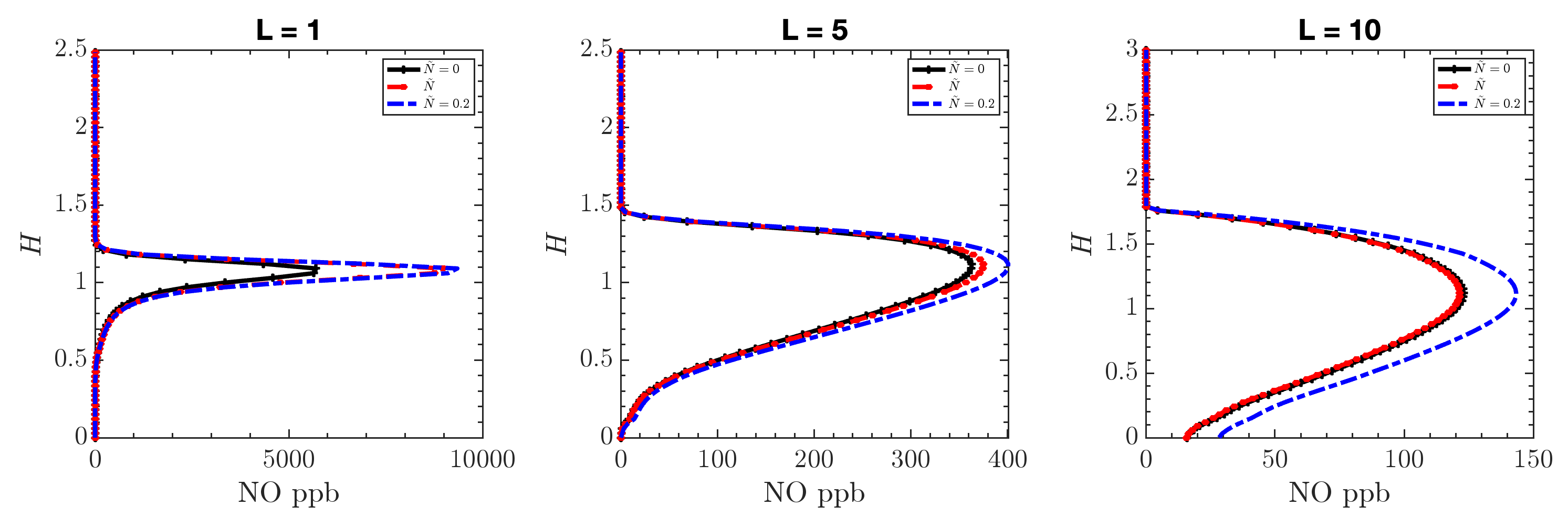}
     \end{overpic}\\
             \begin{overpic}[scale=0.25]{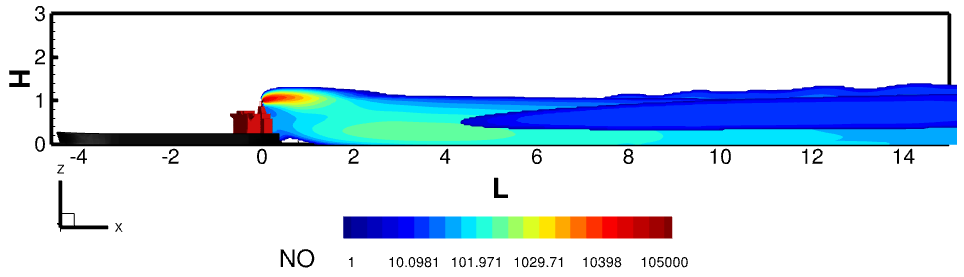}
     \end{overpic} &
         \begin{overpic}[scale=0.2]{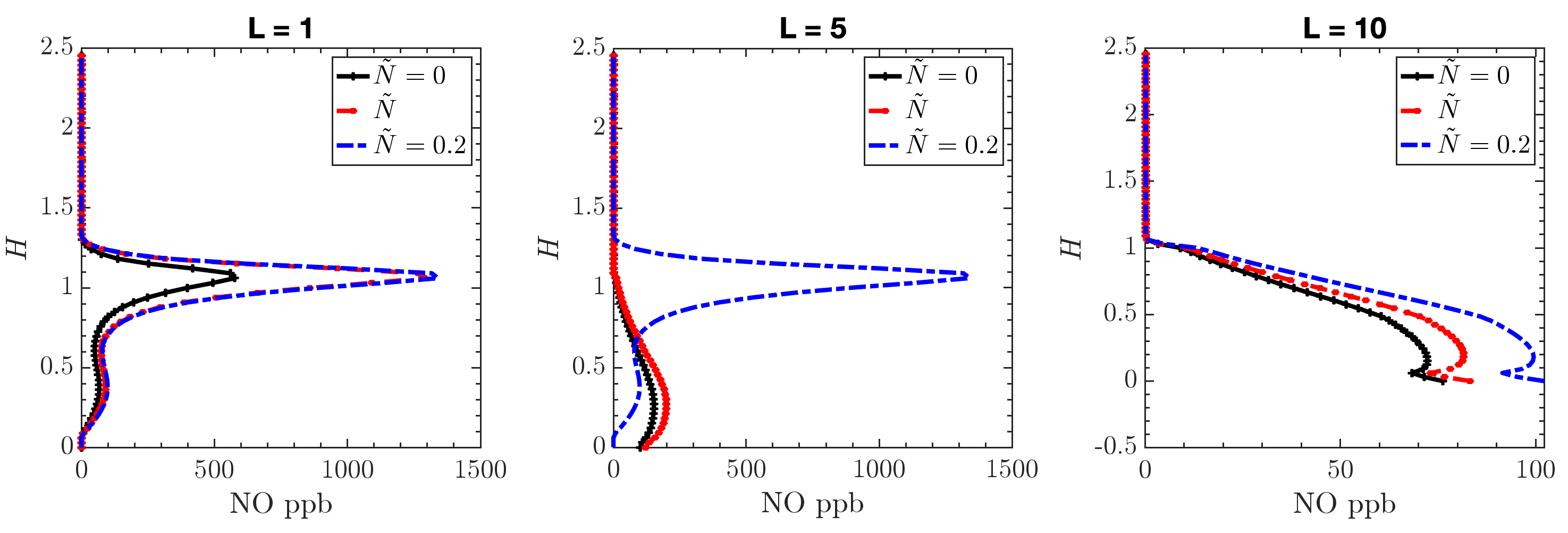}
     \end{overpic}\\
      \begin{overpic}[scale=0.25]{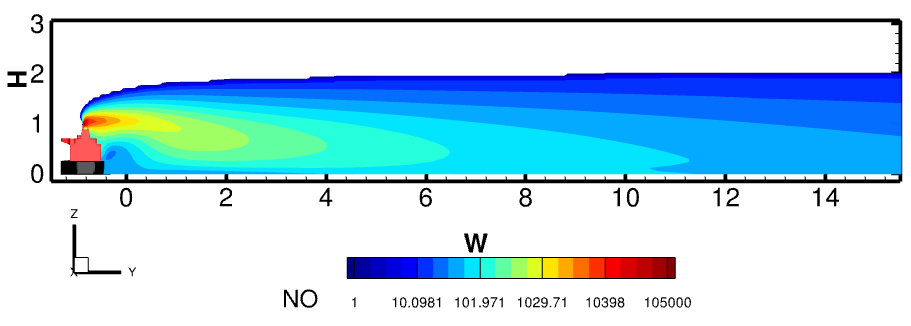}
     \end{overpic} & \begin{overpic}[scale=0.2]{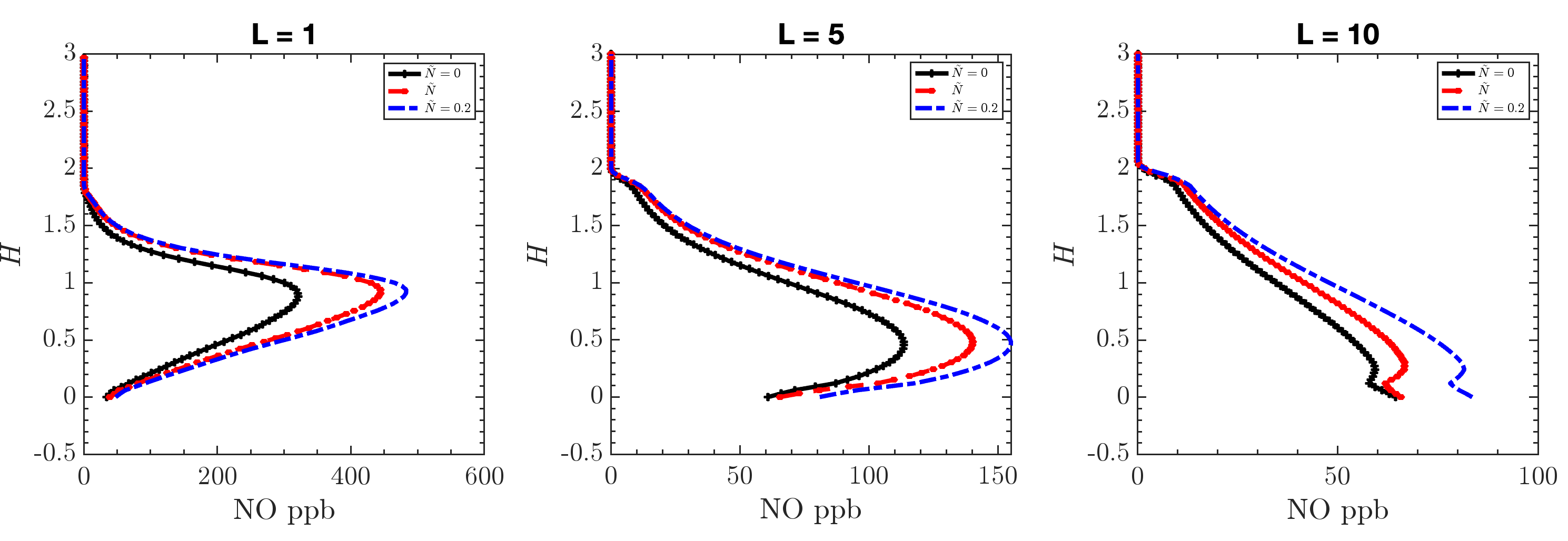}
     \end{overpic}\\
\end{tabular}
     \end{adjustbox}
     \caption{Unconditionally-averaged $NO$ concentrations. Left column: contours on a $x-z$ plane at $y=0$  with $\tilde{N}$. Right column: Line plots comparing the unconditional 
 values  with naturally evolving scalar dissipation Rate $\tilde N$, $\tilde N =0 $ and $\tilde N = 0.2$ at $L=1,5,10$. Upper row for headwind, middle row for inclined wind, and lower row for crosswind cases.}
     \label{fig:UnconditionalAverages_NO}
    \end{figure}

The evolution of PAN, formed photochemically from NO$_x$ and VOC precursors, is consistent across both frameworks: micromixing accelerates early-time production near the wake but its relative influence declines downstream as dilution and chemistry dominate.
Figure \ref{fig:UnconditionalAverages_NO} displays the unconditional averaged concentration of NO at $y=0$ with scalar dissipation rates $\tilde{N} =0, \tilde{N}=0.2 \text{ and } \tilde{N}$ as given in \eqref{eqn:sdr} and at $L =1.0, L =5 \text{ and } L =10$ for the three scenarios: the headwind,  inclined wind and crosswind cases. These figure highlights the influence of the wake region on NO concentration distribution and the effects of micromixing as the species evolve. Notably, the crosswind and inclined wind scenarios exhibit a more significant downwash effect compared to the headwind case. This is likely due to stronger mixing within the wake region for the crosswind and inclined wind cases. The inclined wind, in particular, likely induces a more substantial downward flow of air behind the object, enhancing pollutant mixing in the wake region.
Figure S9 in the Supplementary Material reports the PAN distributions and confirms that micromixing effects are localized for that species, with modest downstream influence relative to NO.


\section{Conclusions}

This study quantifies, for the first time in a maritime context, the role of turbulence-driven micromixing in shaping the
chemical evolution of ship exhaust plumes within the near field. By combining high-fidelity CFD with CMC and ISRN
methods, both macromixing and micromixing effects are resolved in space and mixture-fraction coordinates, offering
insights not attainable with traditional Gaussian plume approaches. The results are relevant for emissions entering coastal boundary layers and for port or slow-sailing operations where nearby receptors exist.

Key findings are:
\begin{itemize}
    \item Micromixing alters NO, NO$_2$, O$_3$, and PAN, with conditional analyses indicating NO reductions of roughly 10--20\% and PAN increases of around 15\% at higher scalar dissipation rates; effects taper downstream but remain species dependent.
    \item Gaussian plume models can overestimate NO by up to 45\% in the first few hundred metres under complex wake--wind interactions; these comparisons provide contextual validation rather than the primary contribution.
    \item Wake-induced downwash and asymmetry in inclined winds push the plume downward toward deck and sea level, underscoring regulatory relevance near ports and coastal corridors.
\end{itemize}

Unlike previous CFD plume studies, this work explicitly resolves micromixing--chemistry coupling, offering a mechanistic foundation for subgrid parameterizations in air-quality modeling frameworks. Prior CFD plume studies such as \citet{boikos2024validating} and \citet{orlandi2024air} focus primarily on macromixing processes or integrated forecasting without explicit turbulence--chemistry coupling; the coupled CFD--CMC--ISRN framework here quantifies small-scale interactions and their influence on reactive species, offering a physically consistent contribution to near-field maritime emission modelling.

The framework developed here provides physically consistent reference data for testing parameterisations of micromixing effects in next-generation dispersion and air-quality models. Future work should extend this approach to aerosol processes (sulfates, black/brown carbon) and validate against in-situ ship plume measurements.

\section*{Credit authorship contribution statement}
\begin{itemize}
\item \textbf{Ramesh Kolluru}:  Execution, Methodology, Software, Formal analysis, Writing -- original draft.
\item \textbf{Savvas Gkantonas}: Conceptualization, Methodology, Writing -- review \& editing.
\item \textbf{B. Harikrishnan}: Software, Visualization.
\item \textbf{Epaminondas Mastorakos}: Supervision, Conceptualization, Writing -- review \& editing, Funding acquisition.
\end{itemize}
\section*{Acknowledgements}
This research is supported by the National Research Foundation, Prime Minister's Office, Singapore, under its Campus for Research Excellence and Technological Enterprise (CREATE) program. We thank Mr.~Mark Pearson from Maran Tankers Management Inc. for supplying information on the tanker modelled here and for very fruitful discussions on marine pollution.

\section*{Declaration of Competing Interest}
The authors declare that they have no known competing financial interests or personal relationships that could have appeared to influence the work reported in this paper.

\section*{Use of AI-assisted tools}
No generative AI or AI-assisted technologies were used in the preparation of this manuscript; writing, analysis, and figure generation relied solely on conventional research tools and author expertise.

 \medskip
 
 \bibliography{references}

@article{mastorakos2003conditional,
  title={The conditional moment closure approach for atmospheric pollution problems},
  author={Mastorakos, E},
	journal={WIT Transactions on Ecology and the Environment},
  volume={66},
  year={2003},
  publisher={WIT Press}
}

@inproceedings{gkantonas_ISRN_2020,
	location = {Orlando, {FL}},
	title = {Incompletely Stirred Reactor Network Modeling of a Model Gas Turbine Combustor},
	isbn = {978-1-62410-595-1},
	doi = {https://doi.org/10.2514/6.2020-2087},
	eventtitle = {{AIAA} Scitech 2020 Forum},
	booktitle = {{AIAA} Scitech 2020 Forum},
	publisher = {American Institute of Aeronautics and Astronautics},
	author = {Gkantonas, Savvas and Giusti, Andrea and Mastorakos, Epaminondas},
	urdate ={2022-11-22},
	year ={2020-01-06},
	langid = {english}
}

@article{gery_photochemical_1989,
	title = {A photochemical kinetics mechanism for urban and regional scale computer modeling},
	volume = {94},
	issn = {0148-0227},
	doi = {https://doi.org/10.1029/JD094iD10p12925},
	pages = {12925},
	issue = {D10},
	journaltitle = {Journal of Geophysical Research},
	shortjournal = {J. Geophys. Res.},
	author = {Gery, Michael W. and Whitten, Gary Z. and Killus, James P. and Dodge, Marcia C.},
	urldate = {2022-12-01},
	date = {1989},
	langid = {english}
}

@article{gkantonas_0d-cmc_nodate,
	title = {0D-{CMC} for calculation of ignition, extinction and fully burning flamelets with detailed chemistry: description of the code and implementation},
	pages = {10},
	author = {Gkantonas, Savvas},
	langid = {english}
}

@article{gkantonas_incompletely_2019,
	title = {Incompletely stirred reactor network modelling for soot emissions prediction in aero-engine combustors},
	pages = {6},
    journaltitle={nternational Workshop on Clean Combustion: Principles and Applications Darmstadt, Germany
},
	author = {Gkantonas, S and Giusti, A and Mastorakos, E},
	date = {2019},
	langid = {english}
}

@article{NGUYEN2022104724,
title = {Reduced order models for uncertainty quantification of gas plumes from leakages during LNG bunkering},
journal = {Journal of Loss Prevention in the Process Industries},
volume = {76},
pages = {104724},
year = {2022},
issn = {0950-4230},
doi = {https://doi.org/10.1016/j.jlp.2022.104724},
author = {Vinh-Tan Nguyen and Venugopalan S.G. Raghavan and Raymond Y.L. Quek and Lim Boon How and Deguang Yan}
}

@article{kim_investigation_2009,
	title = {Investigation of ship-plume chemistry using a newly-developed photochemical/dynamic ship-plume model},
	volume = {9},
	issn = {1680-7324},
	doi = {https://doi.org/10.5194/acp-9-7531-2009},
	pages = {7531--7550},
	number = {19},
	journaltitle = {Atmospheric Chemistry and Physics},
	shortjournal = {Atmos. Chem. Phys.},
	author = {Kim, H. S. and Song, C. H. and Park, R. S. and Huey, G. and Ryu, J. Y.},
	urldate = {2022-12-22},
	date = {2009-10-09},
	langid = {english}
}

@article{garmory_numerical_2015,
	title = {Numerical simulation of oxy-fuel jet flames using unstructured {LES}-{CMC}},
	volume = {35},
	issn = {1540-7489},
	doi = {https://doi.org/10.1016/j.proci.2014.05.032},
	pages = {1207--1214},
	number = {2},
	journaltitle = {Proceedings of the Combustion Institute},
	shortjournal = {Proceedings of the Combustion Institute},
	author = {Garmory, A. and Mastorakos, E.},
	urldate = {2023-01-04},
	year = {2015},
	langid = {english}
}

@article{Park_2016,
author = {Sunho Park, Jiman Yang and Shin Hyung Rhee},
title = {Parametric study on ships exhaust-gas behavior using computational fluid dynamics},
journal = {Engineering Applications of Computational Fluid Mechanics},
volume = {11},
number = {1},
pages = {159-171},
year = {2017},
publisher = {Taylor & Francis},
doi = {https://doi.org/10.1080/19942060.2016.1260057}
}

@article{eyring_2010,
   author = {Veronika Eyring and Ivar S A Isaksen and Terje Berntsen and William J Collins and James J Corbett and Oyvind Endresen and Roy G Grainger and Jana Moldanova and Hans
Schlager and David S Stevenson},
   doi = {https://doi.org/10.1016/j.atmosenv.2009.04.059},
   issn = {1352-2310},
   issue = {37},
   journal = {Atmospheric Environment},
   year = {2010},
   pages = {4735--4771},
   title = {Transport impacts on atmosphere and climate: Shipping},
   volume = {44}
}

@article{Lindstad_2015,
   author = {Haakon Lindstad and Gunnar S. Eskeland and Harilaos Psaraftis and Inge Sandaas and Anders H. Str\o{}mman},
   issn = {00298018},
   journal = {Ocean Engineering},
   month = {12},
   pages = {94--101},
   title = {Maritime shipping and emissions: A three-layered, damage-based approach},
   volume = {110},
   year = {2015},
   doi = {https://doi.org/10.1016/j.oceaneng.2015.09.029}
}

@article{Franke_2008,
   author = {Klaus Franke and Veronika Eyring and Rolf Sander and Johannes Hendricks and Axel Lauer and Robert Sausen},
   doi = {https://doi.org/10.1127/0941-2948/2008/0277},
   issn = {0941-2948},
   issue = {2},
   journal = {Meteorologische Zeitschrift},
   month = {4},
   pages = {117--129},
   title = {Toward effective emissions of ships in global models},
   volume = {17},
   year = {2008}
}

@inproceedings{Garmory_2010,
   author = {Andrew Garmory and Epaminondas Mastorakos and Robert Bilger},
   doi = {https://doi.org/10.2514/6.2010-429},
   journal = {48th AIAA Aerospace Sciences Meeting Including the New Horizons Forum and Aerospace Exposition},
   publisher = {American Institute of Aeronautics and Astronautics},
   title = {Simulations of the Chemical Transformations In a Jet Engine Exhaust Plume}
}

@article{DINATALE2015166,
title = {Particulate matter in marine diesel engines exhausts: Emissions and control strategies},
journal = {Transportation Research Part D: Transport and Environment},
volume = {40},
pages = {166--191},
year = {2015},
issn = {1361-9209},
doi = {https://doi.org/10.1016/j.trd.2015.08.011},
author = {Francesco {Di Natale} and Claudia Carotenuto}
}

@article{Pan_2021,
AUTHOR = {Pan, K. and Lim, M. Q. and Kraft, M. and Mastorakos, E.},
TITLE = {Development of a moving point source model for shipping emission dispersion modeling in EPISODE--CityChem v1.3},
JOURNAL = {Geoscientific Model Development},
VOLUME = {14},
YEAR = {2021},
NUMBER = {7},
PAGES = {4509--4534},
DOI = {https://doi.org/10.5194/gmd-14-4509-2021}
}

@article{Briggs_1973,
   author = {Gary A. Briggs},
   journal = {Contribution No.79,Atmospheric Turbulence and Diffusion Laboratory 1973 Annual Report},
   title = {Diffusion estimation of small emissions},
   year = {1973}
}

@article{HOLMES_2006,
title = {A review of dispersion modelling and its application to the dispersion of particles: An overview of different dispersion models available},
journal = {Atmospheric Environment},
volume = {40},
number = {30},
pages = {5902-5928},
year = {2006},
issn = {1352-2310},
doi = {https://doi.org/10.1016/j.atmosenv.2006.06.003},
author = {N.S. Holmes and L. Morawska}
}

@article{pasquill1979atmospheric,
  title={Atmospheric dispersion modeling},
  author={Pasquill, Frank},
  journal={Journal of the Air Pollution Control Association},
  volume={29},
  number={2},
  pages={117--119},
  year={1979},
  publisher={Taylor \& Francis}
}

@article{Eurocode1,
author = {Gulvanessian, H.},
title = {EN1991 Eurocode 1: Actions on structures},
journal = {Proceedings of the Institution of Civil Engineers - Civil Engineering},
volume = {144},
number = {6},
pages = {14-22},
year = {2001},
doi = {https://doi.org/10.1680/cien.2001.144.6.14}
}

@article{von_glasow_modeling_2003,
   author = {R von Glasow and M G Lawrence and R Sander and P J Crutzen},
   doi = {https://doi.org/10.5194/acp-3-233-2003},
   issue = {1},
   pages = {233-250},
   title = {Modeling the chemical effects of ship exhaust in the cloud-free marine boundary layer},
   volume = {3}
}

@article{Green_1980,
author = {A. E. S. Green, R. P. Singhal and R. Venkateswar},
title = {Analytic Extensions of the Gaussian Plume Model},
journal = {Journal of the Air Pollution Control Association},
volume = {30},
number = {7},
pages = {773-776},
year = {1980},
publisher = {Taylor & Francis},
doi = {https://doi.org/10.1080/00022470.1980.10465108}
}

@article{yang2020modification,
  title={Modification and validation of the Gaussian plume model (GPM) to predict ammonia and particulate matter dispersion},
  author={Yang, Zijiang and Yao, Qi and Buser, Michael D and Alfieri, Joseph G and Li, Hong and Torrents, Alba and McConnell, Laura L and Downey, Peter M and Hapeman, Cathleen J},
  journal={Atmospheric Pollution Research},
  volume={11},
  number={7},
  pages={1063--1072},
  year={2020},
  publisher={Elsevier}
}

@article{boikos2024validating,
  title={Validating CFD modelling of ship plume dispersion in an urban environment with pollutant concentration measurements},
  author={Boikos, Christos and Siamidis, Panagiotis and Oppo, Sonia and Armengaud, Alexandre and Tsegas, George and Mellqvist, Johan and Conde, Vladimir and Ntziachristos, Leonidas},
  journal={Atmospheric Environment},
  volume={319},
  pages={120261},
  year={2024},
  publisher={Elsevier}
}

@article{mamarikas_assessing_2023,
	title = {Assessing Shipping Induced Emissions Impact on Air Quality with Various Techniques: Initial Results of the {SCIPPER} project},
	volume = {72},
	issn = {2352-1465},
	doi = {https://doi.org/10.1016/j.trpro.2023.11.699},
	pages = {2141--2148},
	journaltitle = {Transportation Research Procedia},
	author = {Mamarikas, Sokratis and Matthias, Volker and Karl, Matthias and Fink, Lea and Simonen, Pauli and Keskinen, Jorma and Maso, Miikka Dal and Fridell, Erik and Moldanova, Jana and Hallquist, \AA{}sa and Mellqvist, Johan and Conde, Vladimir and Verbeek, Ruud and Duyzer, Jan and Dinther, Dani\"{e}lle van and Timonen, Hilkka and Jalkanen, Jukka-Pekka and Sundstr{\"o}m, Anu-Maija and Majam{\"a}ki, Elisa and Stylogiannis, Antonios and Ntziachristos, Vasilis and Smyth, Tim and Yang, Mingxi and Deakin, Anthony and Proud, Richard and Oeffner, Johannes and Weisheit, Jonathan and Beecken, J{\"o}rg and Weigelt, Andreas and Griesel, Simone and Schoppmann, Hannes and Oppo, Sonia and Armengaud, Alexandre and D'Anna, Barbara and Temine-Roussel, Brice and Lanzafame, Grazia-Maria and Knudsen, Bettina and Knudsen, Jon and Kuosa, Maria and Irjala, Matti and Buckers, Leo and Vliet, Jasper van and Ntziachristos, Leonidas},
	year = {2023}
}

@article{report_D31,
    title={THE SCIPPER PROJECT, Current status of best practices on ship plume modelling},
    author = {Volker Matthias and Markus Quante and Marvin Lauenburg and Jana Moldanova and Fotios Barmpas and Nikos Rapkos and Leonidas Ntziachristos},
    pages={1--26},
    year={2020} 
}

@article{Song_2003,
author = {Song, C. H. and Chen, G. and Hanna, S. R. and Crawford, J. and Davis, D. D.},
title = {Dispersion and chemical evolution of ship plumes in the marine boundary layer: Investigation of O3/NOy/HOx chemistry},
journal = {Journal of Geophysical Research: Atmospheres},
volume = {108},
number = {D4},
pages = {},
doi = {https://doi.org/10.1029/2002JD002216},
year = {2003}
}

@article{murena2018using,
  title={Using a CFD model to assess the impact of cruise ship emissions on the fa{\c{c}}ades of waterfront buildings in Naples, Italy},
  author={MURENA, FABIO and MELE, BENEDETTO and TOSCANO, DOMENICO and others},
  journal={WIT Trans. Ecol. Environ.},
  volume={230},
  pages={189--197},
  year={2018}
}

@misc{richards2016converge,
	title = {Converge 3.1; Convergent Science: Madison, {WI}, {USA}},
	author = {Richards, K. J. and Senecal, P. K. and Pomraning, E.},
	year = {2023}
}

@techreport{Lee_2017, 
title={Engine international air pollution prevention certificate}, 
number={IMO Number 9779381}, journal={Lloyds Register Asia}, 
institution={Busan Technical Support Office }, 
author={J H Lee (Surveyor to Lloyds Register Asia)}, 
year={2017}, 
month={Jan} 
}

@ARTICLE{OBrien_1991,
       author = {{O'Brien}, Edward E. and {Jiang}, Tai-Lun},
        title = "{The conditional dissipation rate of an initially binary scalar in homogeneous turbulence}",
      journal = {Physics of Fluids A},
         year = 1991,
        month = dec,
       volume = {3},
       number = {12},
        pages = {3121-3123},
          doi = {https://doi.org/10.1063/1.858127}
}

@article{BOIKOS2024111594,
title = {Factors affecting pedestrian-level ship pollution in port areas: CFD in the service of policy-making},
journal = {Building and Environment},
volume = {258},
pages = {111594},
year = {2024},
issn = {0360-1323},
doi = {https://doi.org/10.1016/j.buildenv.2024.111594},
author = {Christos Boikos and Nikolaos Rapkos and Giannis Ioannidis and Sonia Oppo and Alexandre Armengaud and Panagiotis Siamidis and George Tsegas and Leonidas Ntziachristos}
}

@article{orlandi2024air,
  title = {Air quality forecasting of along-route ship emissions using metocean prediction and ship performance models},
  author = {Orlandi, G. and Martinez, L. and Neri, A. and Montanari, A. and Gualtieri, C. and Bartocci, P.},
  journal = {Ocean Engineering},
  volume = {293},
  pages = {116644},
  year = {2024},
  issn = {0029-8018},
  doi = {https://doi.org/10.1016/j.oceaneng.2023.116644},
  url = {https://doi.org/10.1016/j.oceaneng.2023.116644},
  publisher = {Elsevier}
}
\end{document}